\documentclass[10pt]{iopart}

\usepackage{iopams}  
\usepackage{graphicx}
\usepackage{soul} %strickout 
\usepackage{color}
\usepackage{cuted}
\usepackage{float}
\usepackage{bibunits}

\defaultbibliography{Rouxinol_HybridQuantumSystem_final}
\defaultbibliographystyle{iopart-num}

\newcommand{\bra}[1]{\ensuremath{\left\langle#1\right|}}
\newcommand{\ket}[1]{\ensuremath{\left|#1\right\rangle}}

\begin{document}

\title{Measurements of nanoresonator-qubit interactions in a hybrid quantum electromechanical system}
%latex iop
\author{F. Rouxinol$^1$\footnote{Present address:Institute of Physics ``Gleb Wataghin", University of Campinas - UNICAMP, 13083-859, Campinas, S\~{a}o Paulo, Brazil}, Y. Hao$^1$, F. Brito$^2$, A.O. Caldeira$^3$, E.K. Irish$^4$ and M.D. LaHaye$^1$}

\address{$^1$  Department of Physics, Syracuse University, Syracuse NY 13244-1130 USA \newline
$^2$ Instituto de F\'{i}sica de S\~{a}o Carlos, Universidade de S\~{a}o Paulo, C.P. 369, 13560-970, S\~{a}o Carlos, SP Brazil\newline
$^3$ Institute of Physics "Gleb Wataghin", University of Campinas - UNICAMP, 13083-859, Campinas, S\~{a}o Paulo, Brazil\newline
 $^4$ Physics and Astronomy, University of Southampton, Highfield, Southampton SO17 1BJ, United Kingdom}

\ead{mlahaye@syr.edu}
\vspace{10pt}
\begin{indented}
\item[]\today
\end{indented}

\begin{abstract}
Experiments to probe the basic quantum properties of motional degrees of freedom of mechanical systems have developed rapidly over the last decade. One promising approach is to use hybrid electromechanical systems incorporating superconducting qubits and microwave circuitry. However, a critical challenge facing the development of these systems is to achieve strong coupling between mechanics and qubits while simultaneously reducing coupling of both the qubit and mechanical mode to the environment. Here we report measurements of a qubit-coupled mechanical resonator system consisting of an ultra-high-frequency nanoresonator and a long coherence-time superconducting transmon qubit, embedded in superconducting coplanar waveguide cavity. It is demonstrated that the nanoresonator and transmon have commensurate energies and transmon coherence times are one order of magnitude larger than for all previously reported qubit-coupled nanoresonators. Moreover, we show that numerical simulations of this new hybrid quantum system are in good agreement with spectroscopic measurements and suggest that the nanoresonator in our device resides at low thermal occupation number, near its ground state, acting as a dissipative bath seen by the qubit. We also outline how this system could soon be developed as a platform for implementing more advanced experiments with direct relevance to quantum information processing and quantum thermodynamics, including the study of nanoresonator quantum noise properties, reservoir engineering, and nanomechanical quantum state generation and detection.
\end{abstract}

\vspace{2pc}
\noindent{\it Keywords}: hybrid quantum systems, nanomechanics, superconducting qubits 
\maketitle
\ioptwocol
\begin{bibunit}
\section{Introduction}
Experiments over the last decade have begun to demonstrate basic quantum properties of the motional degrees of freedom of mechanical systems at the nano- and microscale\cite{o2010quantum,aspelmeyer2012quantum,safavi2012observation,palomaki2013entangling,lecocq2015resolving,wilson2015measurement,wollman2015quantum,lecocq2015quantum,pirkkalainen2015squeezing}.  These developments have been driven by the prospects of utilizing such mechanical elements both for the fundamental exploration of quantum physics in new \textit{macroscopic} limits\cite{aspelmeyer2012quantum} and for a range of applications including quantum information processing\cite{kurizki2015quantum} and closely-related topics in quantum-assisted sensing and simulation\cite{xiang2013hybrid}. Indeed, with mechanical quantum systems now being engineered and studied, specific applications have begun to crystallize, such as quantum coherent optical-microwave converters\cite{bochmann2013nanomechanical,andrews2014bidirectional,andrews2014bidirectional}, memory elements\cite{palomaki2013coherent}, and signal processing circuitry\cite{massel2011microwave,kerckhoff2013tunable}.  Moreover, these recent results have also stimulated new ideas for architectures to implement quantum simulators\cite{lozada2015quantum}, quantum-state generation\cite{abdi2015quantum} and studies in quantum thermodynamics\cite{campisi2011colloquium}.   

Quantum electromechanical systems, in which superconducting qubits are integrated with nanomechanical devices, present a potentially powerful platform for all of these applications\cite{o2010quantum,lecocq2015resolving,lahaye2009nanomechanical,pirkkalainen2013hybrid,gustafsson2014propagating}.  In these hybrid systems, a Josephson-junction-based qubit\cite{clarke2008superconducting} provides a strong nonlinearity that could be utilized for engineering and measuring a wide range of non-classical mechanical states\cite{abdi2015quantum,irish2003quantum,clerk2007using,armour2008probing,utami2008entanglement,asadian2014probing}.  It has thus been anticipated that these systems could enable investigations of fundamental aspects of quantum behavior at the nanoscale that are relevant to current research in quantum information, including decoherence\cite{armour2008probing,utami2008entanglement}, entanglement\cite{armour2008probing,utami2008entanglement}, and non-locality\cite{asadian2014probing}. Additionally, because of the compact size of nanomechanical devices, qubit-coupled mechanical elements have strong prospects for integration with superconducting quantum processing architectures for use as quantum circuit elements, such as coherent switches\cite{mariantoni2008two} and quantum memory and bus elements\cite{cleland2004superconducting}.  

However, the experimental realization of qubit-coupled mechanical devices is relatively underdeveloped (in comparison with the more experimentally active area of cavity mechanics\cite{aspelmeyer2014cavity}), with just a handful of results published demonstrating basic interactions between the two systems\cite{o2010quantum,lahaye2009nanomechanical,pirkkalainen2013hybrid,gustafsson2014propagating}.  The central challenge facing further development of qubit-mechanics for more advanced experiments or applications is to achieve strong coupling between mechanical and qubit degrees  of freedom (generically denoted as $\lambda$) while simultaneously engineering weak environmental coupling to both the mechanical mode and qubit (generically denoted as $\kappa$ and $\gamma$ respectively)\cite{lahaye2015superconducting}. In a general sense, $\lambda$ sets the characteristic time scale for the dynamics of the coupled system and thus determines how quickly a state can be measured, prepared, or transferred between devices using the mutual interaction. It is therefore essential to achieve $\lambda>\kappa,\gamma$ (the strong-coupling regime) during a particular operation to ensure high fidelity interaction protocols.  

Here we report and analyze measurements of a qubit-coupled nanomechanical device that consists of an ultra-high-frequency (UHF) nanoresonator\cite{huang2005vhf} and superconducting transmon qubit\cite{koch2007charge} embedded in a circuit QED architecture\cite{devoret2013superconducting}. We demonstrate that the nanoresonator and transmon have commensurate energies and that the transmon coherence times are an order-of-magnitude larger than any previously reported qubit-coupled nanoresonators\cite{o2010quantum,lahaye2009nanomechanical,pirkkalainen2013hybrid}, thus presenting a viable new path toward accessing the strong-coupling regime of qubit-mechanics.  We detail the design, fabrication and measurement of this novel three-body hybrid quantum electromechanical system. Moreover, we show that spectroscopic measurements of its behavior can be modelled using numerical simulations based upon a Lindblad master equation with generalized Jayne-Cummings Hamiltonians for the intra-coupled components. We further discuss how the results are consistent with the nanoresonator mode residing at low thermal occupancy ($k_{b}T/\hbar\omega<1$) and acting as dissipative bath for the transmon when the two systems are tuned near resonance, providing prospects for future use of this system to study the influence of controlled environments on transmon's dynamics.  Finally, we discuss how realistic improvements could be implemented to develop the system for more advanced experiments in quantum measurement and state-generation with the nanoresonator.

\section{Methods}

\subsection{Experimental Device}
Figure 1 displays a schematic (Fig. 1a) and images of the main components of the hybrid quantum electromechanical system that we report on here: a T-filtered superconducting coplanar waveguide (CPW) microwave cavity \cite{hao2014development}(Fig. 1b), a Josephson-junction-based transmon-style qubit \cite{koch2007charge} (Fig. 1c), and a flexural nanomechanical resonator (Fig. 1d). Important parameters characterizing the properties of this system are summarized in Table 1.

%%%%% Figure 1

\begin{figure}[t!]
\centering
\includegraphics[ width=1\columnwidth,keepaspectratio]{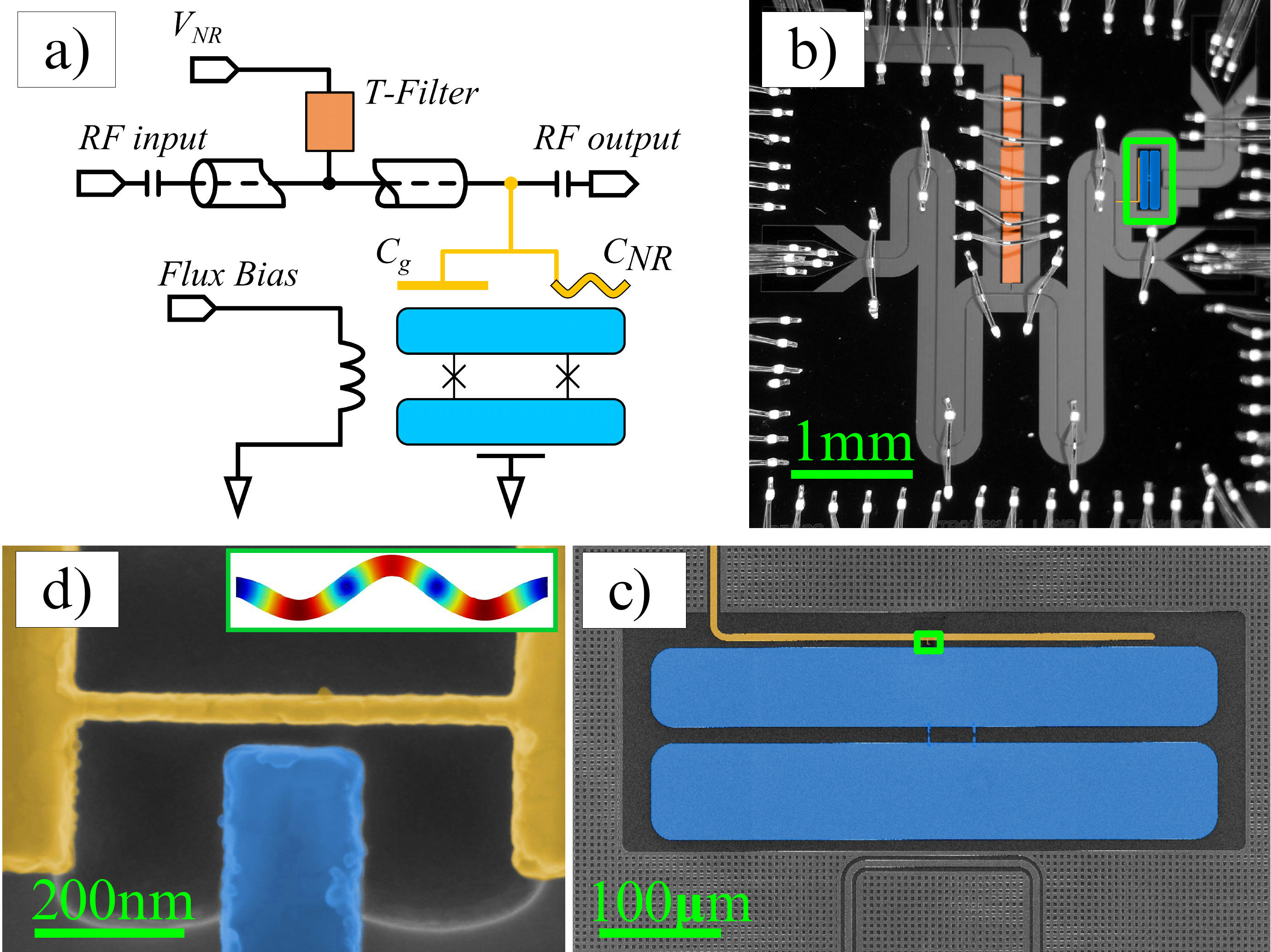}
\caption{ \footnotesize Schematic and images of the hybrid quantum electromechanical device at the focus of this paper. Note that the SEM micrographs are of a device fabricated in the same batch as the one for which data is presented. (a) Circuit schematic of the main components of the device, including the T-filtered CPW cavity, split-junction transmon, and UHF nanoresonator. (b) Optical image of the device, wired for measurement. The CPW and T-filter are of the same design as the device described in Ref. \cite{hao2014development}. (c)  SEM micrograph of region in (b) denoted by the green square. Visible here are the transmon, the CPW stub for coupling the CPW cavity mode and transmon - and for providing $V_{NR}$ between the transmon and nanoresonator - and the superconducting trace for flux biasing the transmon. (d) SEM micrograph of the region in (c) denoted by the green square. The third in-plane mode (inset) of the suspended aluminum nanostructure serves as the nanoresonator in this study. The nanostructure is positioned so that the central anti-node of this mode is located symmetrically about the mid-point of a nearby electrode that extends from the transmon's top shunt capacitor. Due to surface roughness, the separation between the nanostructure and the transmon electrode is observed in the image to vary from $\sim$ 30 nm to 40 nm over the length of the electrode.  
}
    \label{fig1}
\end{figure}

%%%% Table

\begin{table}[t!]
\centering 
\caption{Experimental parameters characterizing the sample. The first set of values shows the CPW and transmon's characteristic energies, and $T_1 $ and $T_2^*$ for $\omega_{01}/2\pi\simeq 4.2 $GHz. The nanomechanical resonator mechanical properties are shown at the second set,  and the last set displays the measured coupling and decoherence rates. Note that the value provided for $\kappa_{cpw}$ is for $V_{NR}=0$.} %title of the table

\begin{tabular}{| c | c |}\hline\hline 
Parameter &Value\\ [1ex]
\hline % inserts single-line
$\omega_{cpw}/2\pi$ & $4.94\mathrm{GHz}$ \\

$E_C/h$&$0.227$ GHz\\ %

$E_{j0}/h$&15.4$ \mathrm{GHz}$ \\

$T_1$ & 15$ \mu\mathrm{s}$ \\

$T_2^*$ & 1.4$ \mu$s \\
\hline\hline

$\omega_{NR}/2\pi$ & 3.4 GHz \\

$\omega_{NR,meas}/2\pi$ & $\simeq 3.47$ GHz \\

\textit{m}&$ 7f\mathrm{g}$ \\

\textit{w} & 45 nm \\

\textit{L}& 700 nm \\

\textit{t} & 100 nm \\

$\rho $ & 2.7 g/cm$^3$\\

$x_{zp}$&$25 \mathrm{fm}$ \\

$\alpha$& 0.447\\

\hline\hline

$g/2\pi$ & 120 $\mathrm{MHz}$\\

$\lambda/(2\pi V_{NR})$ & $\approx 300 \mathrm{kHz/V}$\\

$Q_{NR}$ & 150 \\

$Q_{NR,calc}$ & 280 \\

$\kappa_{cpw}/2\pi$ & 0.28 MHz \\

$\kappa_{NR}/2\pi$ & 24 MHz \\[1ex]

\hline\hline

\end{tabular}
\label{tab1}
\end{table}

At the heart of the device is a suspended aluminum nanostructure (Fig. 1d), with nominal dimensions of 700 nm $\times$ 45 nm $\times$ 100 nm.  Finite element simulations conducted with the commercial software package COMSOL\cite{comsol2011multiphysics} (Fig. 1d, inset) were used to engineer the lowest in-plane flexural modes to have frequencies in the UHF regime. In particular, the third in-plane mode was designed to have a resonant frequency $\omega_{\mathit{NR}}/2\pi = 3.4\,{\rm GHz}$, which was commensurate with the transition energy of the qubit (see below). For the remainder of the text, this mode will be denoted as the nanoresonator. While the nanoresonator's properties were not probed independently of the qubit in this work, microwave spectroscopy of the coupled system (see Section \ref{section:results}) indicated a nanoresonator frequency $\omega_{\it NR,meas}/2\pi = 3.47\,{\rm GHz}$, which is in good agreement with the COMSOL simulations and an estimate of the mechanical quality factor $Q_{\mathit{ NR }} = 150$ ($\kappa_{\mathit{NR}}/2\pi= 24\,{\mathrm{MHz}})$ that is in reasonable agreement with analytical calculations based upon clamping loss through the nanostructure's supports ($Q_{ \mathit{NR},calc}=280$)\cite{wilson2008intrinsic}. 

\begin{figure}[t!]
\centering
\includegraphics[ width=0.9\columnwidth,keepaspectratio]{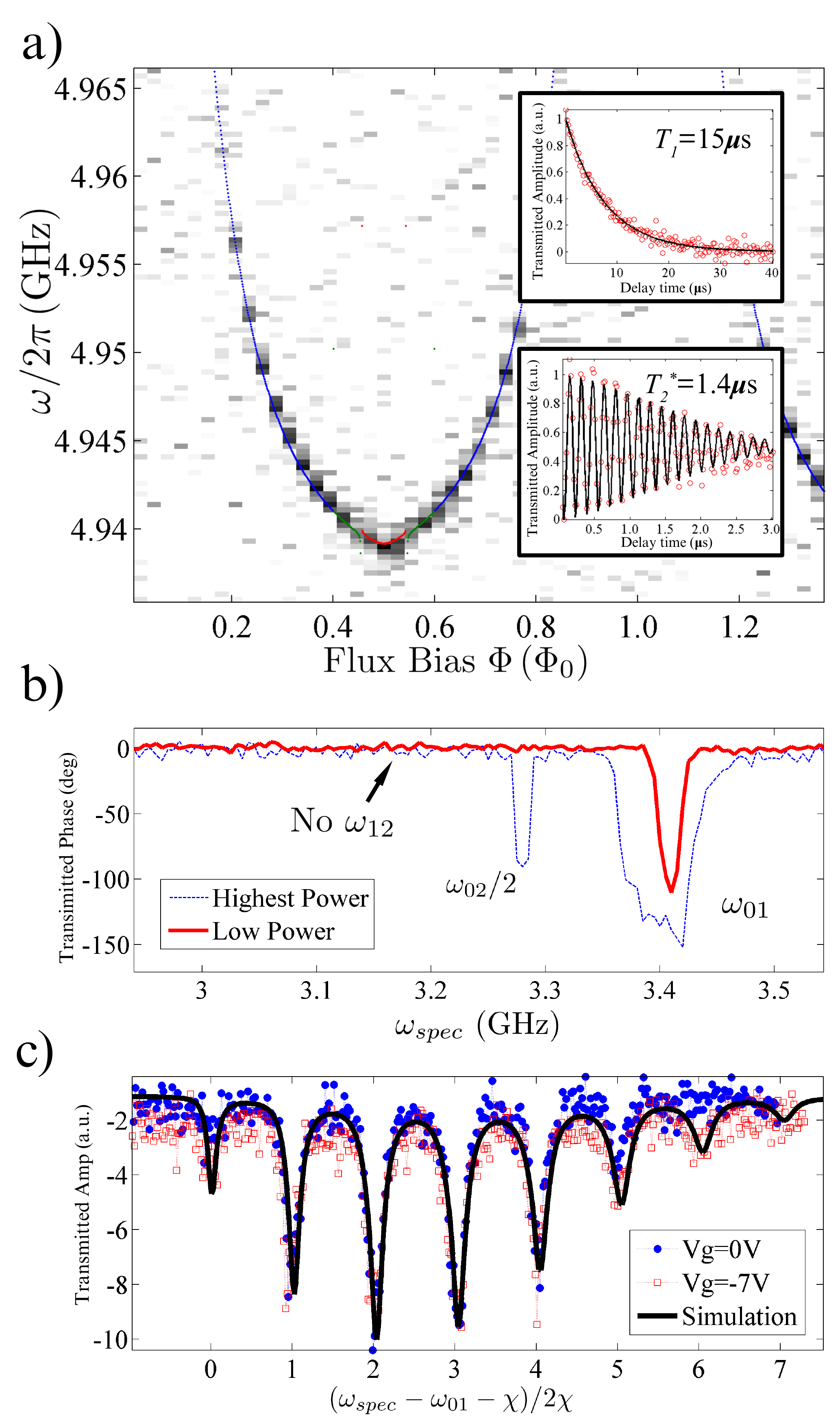}
\caption{ \footnotesize Spectroscopy and time-domain measurements from which the parameter values of the transmon and CPW cavity mode were determined. (a) Single-tone spectroscopy of the CPW with $V_{\mathit{NR}}=0\,{\rm{V}}$, as a function of flux bias $\Phi$ and spectroscopy frequency $\omega$.   Gray scale represents the magnitude of the cavity's transmitted signal. The data  illustrates the dispersive pull of the CPW mode over one flux period $\Phi_0$.  Insets: (top) Measurement of the relaxation time $T_1$ for the transmon's $\omega_{01}$ transition at $\omega_{01}/2\pi=4.2\,{\rm{GHz}}$, where the CPW, transmon and nanoresonator are all far-detuned in energy; (bottom) Ramsey interference measurement of the transmon taken at $\omega_{01}/2\pi=4.2\,{\rm{GHz}}$. (b) Two-tone spectroscopy measurements of the transmon at high and low powers, illustrating the $\omega_{01}$ and $\omega_{02}/2$ transitions. This data was taken at $V_{\mathit{NR}}= -5$ V. No peak is observed at the $\omega_{12}$ transition frequency, indicating that the transmon remains ``cold" when far-detuned from the nanoresonator. (c) Two-tone spectroscopy measurements of the transmon $\omega_{01}$ transition around $\omega_{01}/2\pi=4.2 \rm{GHz}$ for $V_{\mathit{NR}} = 0\,{ \rm{V}}$ and $V_{\mathit{NR}}=-7\,{ \rm{V}}$ as a function of spectroscopy tone $\omega_{spec}$, illustrating the CPW-number-state-resolved Stark shift of the transmon.  The solid line is a numerical simulation including only the transmon and CPW, using $T_1$, $T_2^*$, and $\kappa_{cpw}$ from Table 1, with the transmon and CPW temperatures set to $T_Q=30\,{\rm{mK}}$ and $T_{cpw}=45\,{\rm{mK}}$ (See Supplemental Material).}
    \label{fig2}
\end{figure}

A 2D split-junction transmon served as the qubit in this work (Fig. 1c). The transmon was formed from two Al/AlO/Al Josephson junctions connected in parallel between two large area (200 $\mu$m $\times$ 50 $\mu$m) niobium pads. From spectroscopic measurements (Figs. \ref{fig2}a-b), the charging energy and maximum total Josephson energy of the transmon were determined to be $E_C/h=0.227\,{\rm GHz}$ and $E_{j0}/h=15.4\,{\rm GHz}$. An on-chip niobium flux-bias trace was used to provide magnetic flux $\Phi$ to the transmon in order to tune the transmon's Josephson energy $E_{j}=E_{j0}{\rm }\cos\left( \pi\Phi/\Phi_{0} \right)$ (Fig. 2a). Time-domain measurements of the relaxation time $T_1$ of the transmon's $\omega_{01}$ transition (i.e. $\ket{1}\to \ket{0}$) conducted at a flux bias point $\Phi=0.82\Phi_0$, where $\omega_{01}$ was detuned from both the nanoresonator and CPW fundamental mode (see below), yielded a maximum relaxation time of $T_{1}=15\,\mu$s (top inset, Fig. 2a), which was consistent with estimates of radiative loss through the T-filtered CPW (See Supplemental Material). Ramsey interference measurements made at the same value of $\Phi$ yielded a maximum coherence time of $T_{2}^{*} = 1.4\,\mu$s (bottom inset, Fig. 2a), which we believe was likely limited by dephasing that arises due to the high susceptibility of our symmetric junction design to low frequency flux noise - for the measurements reported here, the transmon was operated in a regime where $\partial{\omega_{01}}/\partial{\Phi}$  is large, leading to reduced $T_2^*$ values in comparison with what is typically observed when the transmon is operated at the ``sweet spot" where $\partial{\omega_{01}}/\partial{\Phi}=0$\cite{koch2007charge}. It is important to point out that these coherence values are more than an order of magnitude larger than achieved in all previously published works\cite{o2010quantum,lahaye2009nanomechanical,pirkkalainen2013hybrid} involving the integration of a nanoresonator and superconducting qubit.

%%%%% Figure 2

The application of a large DC voltage (on the order of Volts) between the nanoresonator and the transmon $V_{\mathit{NR}}$ served to establish coupling between the nanoresonator's flexural motion and the electrostatic energy of transmon. To lowest order, mechanical displacement $x$ of the nanoresonator linearly modulates the transmon's polarization charge through the devices' mutual capacitance $C_{\mathit{NR}}$ (Fig. 1a), resulting in an interaction that is characterized by the coupling strength\cite{irish2003quantum,lahaye2009nanomechanical}

\begin{equation}
\lambda=-4\frac{E_C}{\hbar}\frac{\mathrm{d}C_{\mathit{NR}}}{\mathrm{d}x}\frac{V_{\mathit{NR}}}{e}x_{zp},
\label{eq:couplingNR}
\end{equation}
where $x_{zp}=\sqrt{\hbar/2m\omega_{\mathit{NR}}}$ are the RMS zero-point fluctuations of the mode, with $m=\alpha\rho wLt$ as the effective mass of the resonator, and the parameters $w$, $L$, and $t$ are the geometrical width, length and (out-of-plane) thickness of the structure respectively.  Estimates of these parameters, along with the effective mass ratio factor $\alpha$, are provided in Table \ref{tab1}. From spectroscopic measurements  of the coupled-device (see Section \ref{section:results}), we estimate $\lambda/2\pi V_{\mathit{NR}}\approx 300\, \mathrm{ kHz/V}$, which is the value we use to perform numerical simulations of the coupled-device (Section \ref{section:results}). For this magnitude of coupling, the strong-coupling regime of the transmon and nanoresonator with respect to qubit decoherence (i.e. $\lambda>\frac{2\pi}{T_{2}}$) is accessed for $|V_{\mathit{NR}}|>2 \,{\rm{V}}$. However, the system remains in the weak coupling regime in regard to nanoresonator dissipation, with  $\lambda<\kappa_{\mathit{NR}}$ for all values of $V_{\mathit{NR}}$ explored ($|V_{\mathit{NR}}| \lesssim 8 \,{\rm{V}}$). It should be noted that the value of $\lambda$ extracted from measurements and used in our simulations is within the right range of the values of $\lambda$ predicted by Eq. \ref{eq:couplingNR} using the parameters in Table 1 and the approximation $\frac{\rmd C_{\mathit{NR}}}{\rm{d}x}\simeq\frac{C_{\mathit{NR}}}{d}$, where $d=35\, \mathrm{nm}$ is the spacing between nanostructure and the transmon's coupling electrode; but it exceeds by a factor of 10 estimates based upon simple numerical simulations of $\frac{\rm{d}C_{\mathit{NR}}}{\rm{d}x}$ using finite element simulations (See Supplemental Material).\footnote{$C_{NR}$ was simulated using commercial finite element software. $\rm{d}C_{NR}/\rm{d}x$ was then estimated by calculating $C_{\mathit{NR}}$ for different values of $d$ and numerically calculating $\rm{d}C_{NR}/\rm{d}x$.} The source of the discrepancy remains a subject of ongoing investigations.

In order to isolate and measure the coupled transmon and nanoresonator, these devices were embedded in a circuit QED (cQED) architecture (Fig. 1b)\cite{devoret2013superconducting}. In this configuration, the transmon and nanoresonator were located in a pocket of the ground plane of a superconducting CPW near a voltage anti-node of the CPW's fundamental mode, which had a frequency of $\omega_{cpw}/2\pi =4.94\,{\rm{GHz}}$, and a loaded quality factor $Q_{cpw}=20\times 10^3$ ($\kappa_{cpw}/2\pi = 280\,{\rm kHz}$), when far detuned in energy from the transmon.  A superconducting stub, which extended into the ground plane pocket from the center trace of the CPW, provided capacitive coupling $C_g$ between the CPW's fundamental mode and the transmon. The resulting coupling $g$ was given by the standard expression used in cQED\cite{koch2007charge}
\begin{equation}
g=2\frac{\beta eV_{zp}}{\hbar},
\end{equation}
where $\beta=C_{g}/C_{\Sigma}$ is the ratio of CPW-transmon mutual capacitance $C_{g}$ (Fig. 1a) to the transmon's total capacitance $C_{\Sigma}$, and $V_{zp}=\sqrt{\hbar\omega_{cpw}/2C_{cpw}}
$ are the RMS zero-point fluctuations of the CPW cavity, with total  
capacitance $C_{cpw}$.  For this device, the engineered parameters yielded $g/2\pi\approx$ 120 MHz, providing access to the strong dispersive coupling limit between the transmon and CPW\cite{schuster2007resolving}, where the effective dispersive coupling strength $\chi$ exceeded the linewidths of both the CPW and transmon (i.e. $\chi/2\pi>[\frac{2\pi}{T_{1}},\frac{2\pi}{T_{2}},\kappa_{cpw}]$). This enabled measurements of the number-state-resolved AC Stark shift of transmon's 0-1 transition energy that arises due to the dispersive coupling to the CPW mode (Fig. 2c). Note that the resonant limit between the transmon and CPW cavity was inaccessible because $\omega_{cpw}>\omega_{01}$ for all values of $\Phi$. 
 
The superconducting stub also provided a galvanic connection to the nanoresonator, allowing for the application of DC voltages to supply $V_{\mathit{NR}}$ and establish coupling between the transmon and nanoresonator.  The DC voltage was applied through a superconducting Nb T-filter\cite{hao2014development} that was connected to the mid-point of the CPW cavity and was designed to introduce negligible loss to the CPW in comparison to intrinsic dissipation and losses through the CPW's coupling capacitors $C_{c}$.

%%%%%%%%%%%%%%Fabrication of the Device%%%%%%%%%%%%%%%%%%%%        
\subsection{Fabrication of the Device}

The device was fabricated in a series of steps involving standard micro- and nanolithographic techniques.  First, a 100 nm thick layer of Nb was DC-sputtered on a high-resistivity ($>10\,{\rm k}\Omega\cdot{\rm cm}$) silicon wafer, whose surface was prepared with an ion-mill etch before deposition of the Nb. Next the CPW, T-filter, ground plane, transmon shunt pads, and flux bias trace were patterned from the Nb using deep-UV photolithography followed by a reactive ion etch with gas mixture Ar:BCl$_{3}$:Cl$_{2}$. 

Next the nanoresonator was defined in a lift-off process, using e-beam lithography to define the pattern, and then an aluminum deposition in a dedicated e-beam evaporation system. This was followed by a dry-etch process to free the resonator using a PPMA mask defined by e-beam lithography and a reactive ion plasma of SF$_6$:Ar to undercut the nanostructure. The sample was then cleaned by a soft  oxygen plasma ashing process (descum) to remove any residual resist of the surface of the sample.

In the final step, the transmon's Josephson junctions were fabricated.  This involved a third layer of e-beam lithography, followed by a standard Dolan-bridge double-angle evaporation\cite{fulton1987observation} of aluminum in ultra-high vacuum using an evaporator dedicated to aluminum deposition.

\subsection{Milli-Kelvin Measurement Circuitry and Microwave Spectroscopy}

Measurements of the device were performed at milli-Kelvin temperatures using a dilution refrigerator with a base temperature between 20 and 30 mK.  The devices were enclosed in a light-tight OFHC copper sample holder, which itself was situated in a home-made,  lead-lined Cryoperm magnetic shield; both the sample holder and shield were anchored to the base-stage of the refrigerator. In order to minimize stray radiation and the excitation of nonequilibrium quasiparticles in the superconducting circuitry\cite{barends2011minimizing,corcoles2011protecting}, the inner wall of the sample holder was coated with microwave/infrared absorbing foam and the input and output lines into the sample holder were heavily filtered and attenuated at various stages of the dilution refrigerator as illustrated in the circuit diagram in Figure 3.

\begin{figure}[t!]
\centering
\includegraphics[ width=1\columnwidth,keepaspectratio]{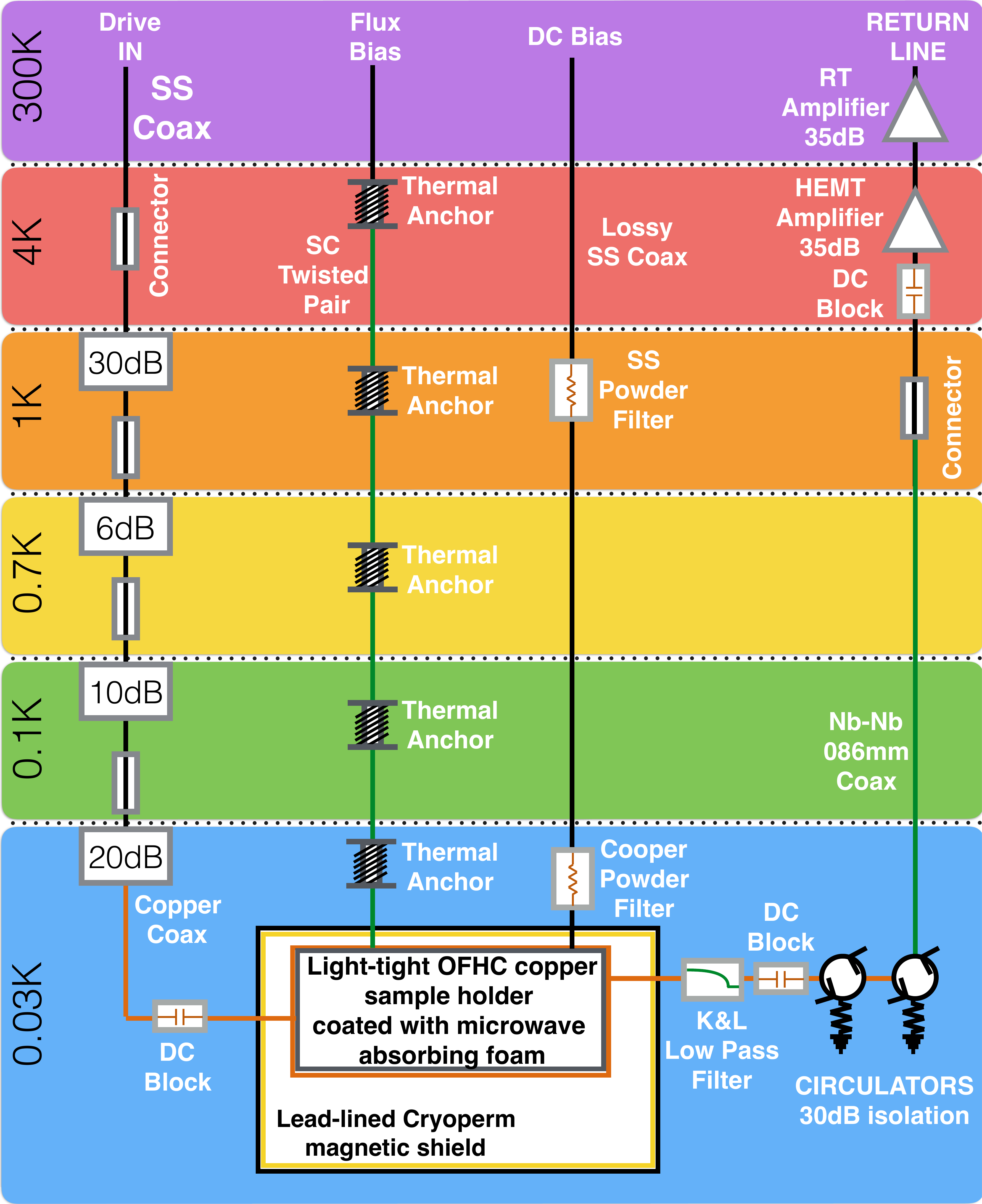}
\caption{ \footnotesize Schematic of measurement circuitry on the dilution refrigerator used for the measurements discussed in Section \ref{section:results}. SS - stainless-steel, SC - Superconductor Niobium-Titanium wire, HEMT - High Electron Mobility Transistor.
}
    \label{fig3}
\end{figure}

The results reported below were obtained from both single-tone and two-tone spectroscopy measurements of the coupled CPW, transmon and nanoresonator. Single-tone spectroscopy was accomplished using pulsed transmission measurements of the CPW at frequencies $\omega$ in the vicinity of $\omega_{cpw}$ for fixed values of $\Phi$ and $V_{\mathit{NR}}$.  The transmitted pulses were measured with a first-stage cryogenic HEMT amplifier, followed by additional room temperature amplifiers and a home-made superheterodyne circuit that mixed the signal to an IF frequency of 10.7 MHz; the IF signal was then digitized with a high-speed ADC, and the phase and amplitude were extracted numerically using a digital homodyne technique.  Measurement pulse lengths of $\sim 1\,{\rm ms}$ were chosen so that the averaging time greatly exceed the relaxation rates of the CPW, transmon and nanoresonator, and thus the data reflected the steady-state behavior of the system. 

Two-tone spectroscopy was performed using two different pulses: first a long ($\sim 40\,{\rm \mu}$s) microwave pulse with variable frequency, $\omega_{spec}$, tuned near the transmon transition frequency $\omega_{01}$, was applied to the cavity to excite the transmon; then a second pulse of 4$\,\mu$s was applied at $\omega_{cpw}$ to perform a dispersive measurement\cite{blais2004cavity} of the transmon's state. The phase and amplitude of the cavity's transmitted signal were then recovered using the same home-made superheterodyne/digital-homodyne setup used for the single-tone spectroscopy. 

\subsection{Model and Numerical Simulations}

In order to simulate the single-tone spectroscopy measurements,\footnote{We did not perform simulations of the two-tone spectroscopy of the complete device, due to the complexity of such simulations, and leave this for future work.} we follow standard approaches for modelling each part of our hybrid system and their respective interactions. Here, the transmon is considered as a multi-level atom\cite{koch2007charge}, represented by the Hamiltonian
\begin{equation}
\hat{H}_{T}=\sum_{m}\hbar\omega_{0m}\ket{m}\bra{m},
\end{equation}
where the eigenenergy differences $\hbar\omega_{0m}$, which are dependent on $E_{j0}$, $E_C$ and the applied magnetic flux are determined using the circuit model theory\cite{burkard2004multilevel,koch2007charge} for the bare device (See Supplemental Material).

The CPW cavity and the nanoresonator are modelled as single mode harmonic oscillators, represented by the canonical bosonic operators $\hat{a}(\hat{a}^\dagger)$ and $\hat{b}(\hat{b}^\dagger)$, respectively. It is worth noting that their natural frequencies $\omega_{cpw}$ and $\omega_{\mathit{NR}}$ are, by design, out of resonance. Moreover their direct coupling is relatively weak.  As a result the CPW-nanoresonator interaction has negligible affect on the unitary evolution of the entire system and thus is omitted from the system Hamiltonian. 

To model the transmon-CPW and transmon-nanoresonator direct couplings, we use for each a multi-level, generalized Jaynes-Cummings Hamiltonian\cite{koch2007charge,pirkkalainen2013hybrid}, 
\begin{eqnarray}
\hat{H}_{T-cpw}=\sum_{l,m}g_{l,m}\ket{l}\bra{m}(\hat{a}^{\dagger}+\hat{a}),\\
\hat{H}_{T-NR}=\sum_{l,m}\lambda_{l,m}\ket{l}\bra{m}(\hat{b}^{\dagger}+\hat{b}),
\end{eqnarray}
with $g_{l,m}=g\bra{l}\hat{n}\ket{m}$ and $\lambda_{l,m}=\lambda\bra{l}\hat{n}\ket{m}$ representing the coupling strength of the $\ket{l} \to  \ket{m}$ transition, where $\hat{n}$ is the Cooper-pair number operator associated with the transmon.

Finally, the microwave field applied to the cavity to perform single-tone spectroscopy is represented by the term
\begin{equation}
\hat{H}_{Drive}=E_{d}(e^{i\omega t}\hat{a}+e^{-i\omega t}\hat{a}^{\dagger}),
\end{equation}
where $\omega$ is the frequency of the signal and $E_{d}$ its amplitude.

We evaluate the system state dynamics by numerically solving the Lindblad form of the system master equation\cite{Weiss} 
\begin{equation}
\frac{d\hat{\rho}}{dt}=-\frac{i}{\hbar}[\hat{H},\hat{\rho}]+\sum_{k}\gamma_k\left(\hat{{\cal A}}_k\rho  \hat{{\cal A}}_k^\dagger-\frac{1}{2}\{\hat{{\cal A}}_k^\dagger \hat{{\cal A}}_k,\rho\}\right),
\end{equation}
where $\hat{\rho}$ represents the hybrid system density matrix and  $\{\hat{{\cal A}}_k^\dagger \hat{{\cal A}}_k,\rho\}\equiv \hat{{\cal A}}_k^\dagger \hat{{\cal A}}_k\rho+\rho \hat{{\cal A}}_k^\dagger \hat{{\cal A}}_k$. The operators $\hat{{\cal A}}_k$ are standard Lindblad operators and are responsible for inducing relaxation and decoherence processes in the state time evolution. We use $(\hat{a}^\dagger,\hat{b}^\dagger,\ket{l+1}\bra{l})$ as the set of Lindblad operators associated with thermally induced absorption processes and $(\hat{a},\hat{b},\ket{l}\bra{l+1})$ as those representing relaxation processes. We include dephasing processes by adding the transmon operators $(\ket{l}\bra{l})$ to our set of Lindblad operators. The respective rates $\gamma_k$ are determined from the knowledge of CPW cavity and nanoresonator quality factors ($\kappa_{cpw}, \kappa_{\mathit{NR}}$), transmon relaxation and decoherence times $(T_1, T_2^\ast)$ and temperature estimations. (See Supplemental Material for more details.) Since we are interested in characterizing quantities in the steady-state regime, we only have to determine the solution of $\dot{\rho}=0$, which we calculate numerically assuming the hybrid system Hilbert space spanned by the three, five and four lowest eigenenergy states of the bare transmon, nanoresonator and CPW cavity, respectively.

\section{Results} 
\label{section:results}
\begin{figure*}[!ht]
\begin{center}\includegraphics[ width=2\columnwidth,keepaspectratio]{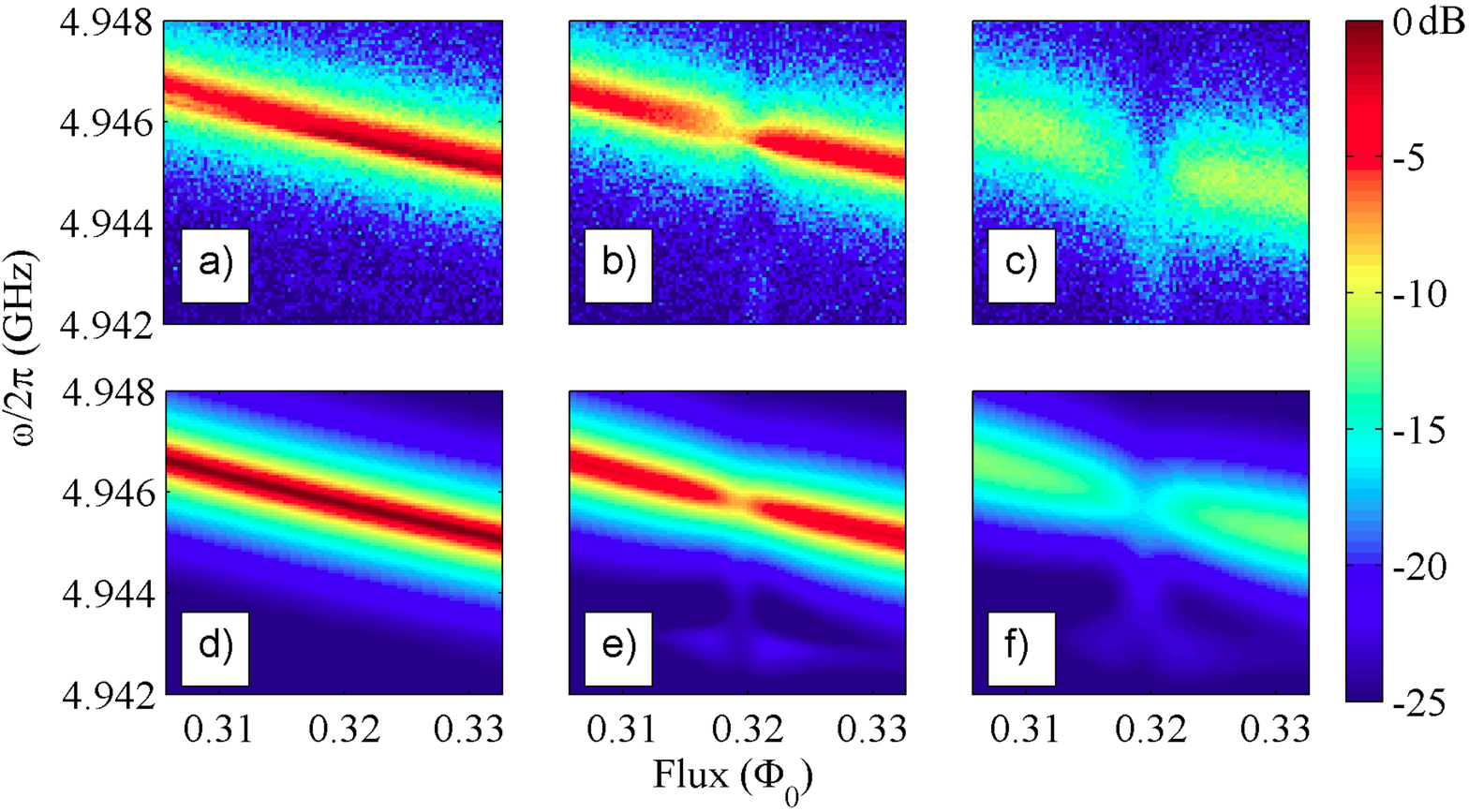}\end{center}
\caption{ \footnotesize Measurements and simulations of single-tone spectroscopy of the cavity mode around $\omega_{cpw}$ as a function of $\Phi$ over a range where $\omega_{01}\approx\omega_{\mathit{NR}}$. The color scale indicates the amplitude of the cavity transmitted signal. Measured data are plotted for three different coupling voltages $V_{\mathit{NR}}=-4.5\,{\rm{V}}, -5.5\,{\rm{V}}, -6.5 \,{\rm{V}}$ (a-c). Simulated data is plotted for $\lambda/2\pi=1.35, 1.65, 1.95\,{ \rm{MHz}}$ (d-f). In the simulations, the nanoresonator temperature $T_{\mathit{NR}}$ is also increased, with $T_{\mathit{NR}}=30\,{\rm {mK}}, 100\,{\rm {mK}}, \rm{and}\, 180\,{\rm {mK}}$ in (d), (e), and (f) respectively. For all three simulation maps, the remaining fit parameters were fixed at the experimentally determined values. (a) Map of single-tone spectroscopy data that is typical of measurements made for $|V_{\mathit{NR}}| \lesssim 5\,{\rm{V}}$, for which the dispersive pull of $\omega_{cpw}$ due to the transmon is apparent, but no manifestation of the nanoresonator-transmon interaction is evident. (b-c) As $\lambda$ increases ($|V_{\mathit{NR}}|> 5\,{\rm{V}}$) and $T_{\mathit{NR}}$ increases due to heating from the leakage current, a gap becomes apparent in the spectroscopy data around $\Phi=0.32 \Phi_0$, where $\omega_{01}=\omega_{\mathit{NR}}$.  The location of this feature was reproducible through repeated cycling of fridge temperature and $V_{\mathit{NR}}$, and was observed to be periodic in $\Phi$ as one would expect given the dependence of $\omega_{01}$ on $E_j(\Phi)$.       
}
    \label{fig4}
\end{figure*}

\begin{figure}[!ht]
\centering
\includegraphics[ width=1\columnwidth,keepaspectratio]{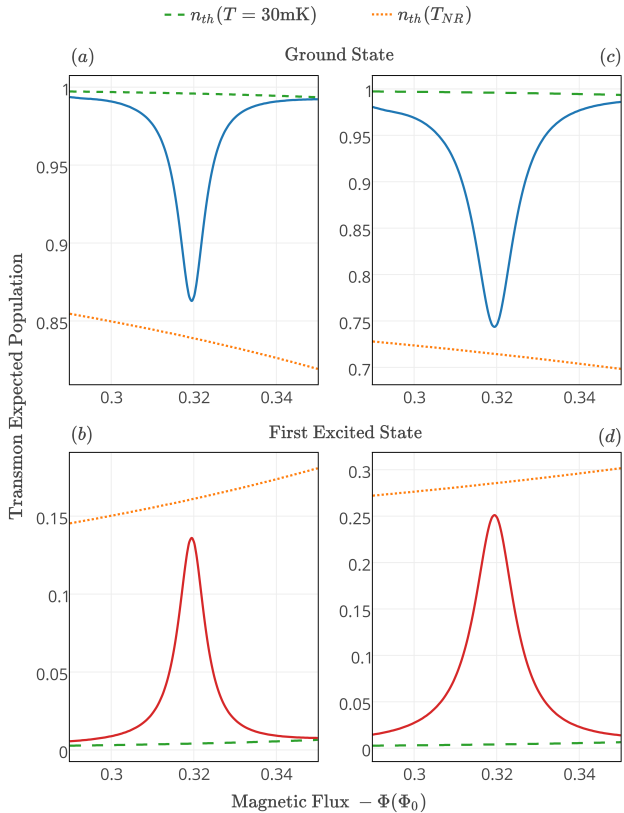}
\caption{\footnotesize Numerical simulations of the expected populations of the transmon ground (panels a and c) and first excited (panels b and d) states as a function of the applied magnetic flux $\Phi$ for the single-tone spectroscopy. The model assumes that the transmon is directly coupled to a $T=30\,{\rm{mK}}$ thermal reservoir.  As the transmon and nanoresonator are almost on resonance $(\Phi\approx 0.32\Phi_0)$, it is observed that the transmon state populations deviate from those imposed by a $T=30\,{\rm{mK}}$ reservoir (green dashed line), to ones much more related with a thermal reservoir at the nanoresonator temperature $T_{NR}$ (orange dotted line). Panels a and b (c and d) show the case $V_{NR}=-5.5\,{\rm{V}}$ $(-6.5\,{\rm{V}})$ for which we estimate the nanoresonator temperatures  $T_{NR}=100\,{\rm{mK}}$ $(180\,{\rm{mK}})$, respectively.}

    \label{fig5}
\end{figure}

\begin{figure}[t!]
\centering
\includegraphics[ width=1\columnwidth,keepaspectratio]{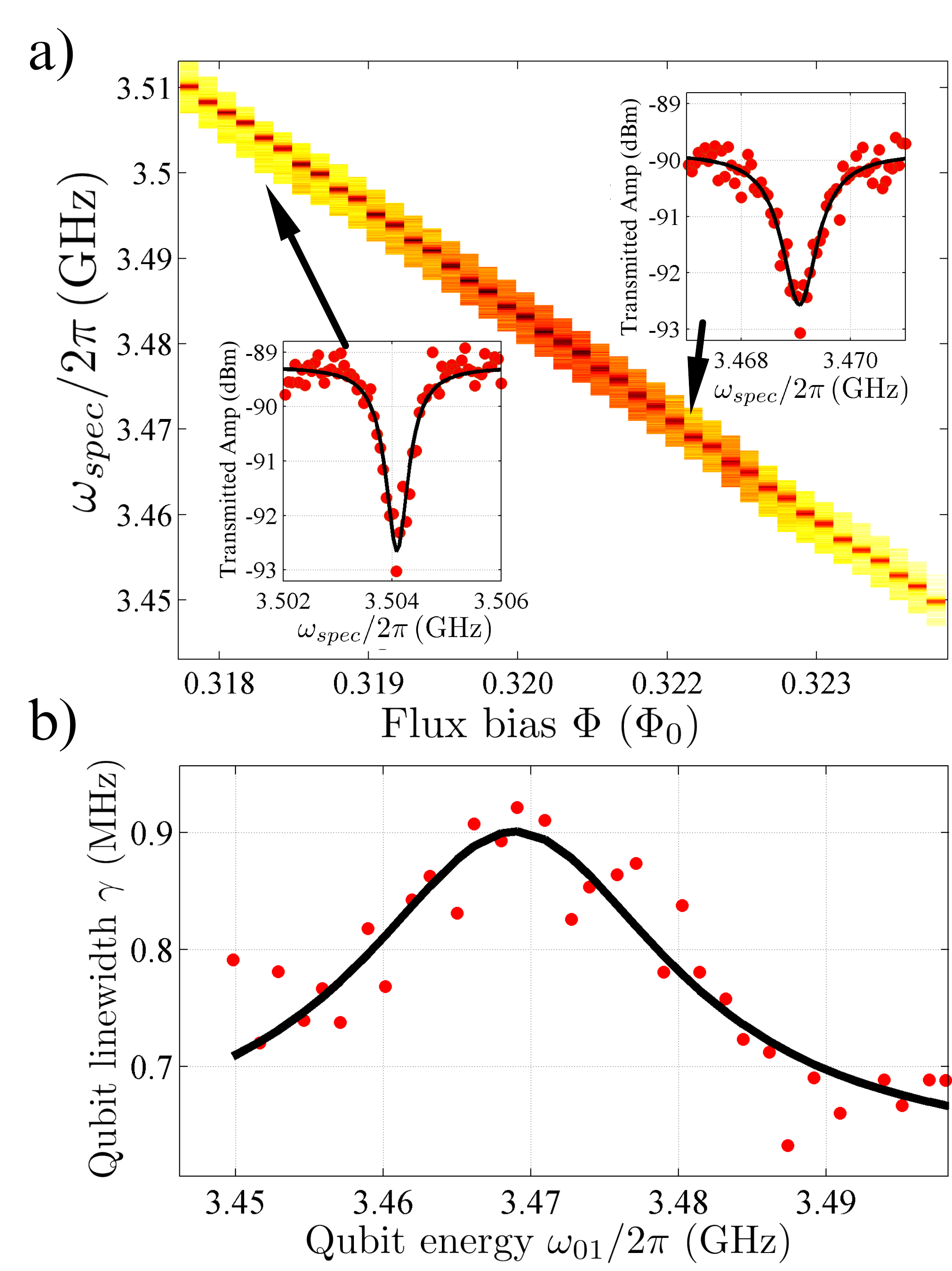}
\caption{ \footnotesize Two-tone spectroscopy measurements of the transmon's $\omega_{01}$ transition for $\Phi$ around $\Phi =0.32\Phi_{0}$, where $\omega_{01}=\omega_{\mathit{NR}}$ for $V_{\mathit{NR}}= -5 V$. For this value of $V_{\mathit{NR}}$, simulations of single-tone spectroscopy (Fig. 4d) indicate that $T_{\mathit{NR}}< 100 \,{\rm{mK}}$ ($n_{th}<0.22$), and thus the nanoresonator should preferentially absorb energy from the transmon. (a) The linewidth of the $\omega_{01}$ transition $\gamma$ exhibits clear broadening in the vicinity of the resonance with the nanoresonator, which a simple model (main text) suggests is proportional to the positive frequency component of the nanoresonator's displacement spectral density $S_{x}(\omega)$  (Eq. \ref{eq:simplegamma}). (b) A fit of $S_x(\omega)$ (solid line) to $\gamma$ over this frequency range allows for the estimates of $\lambda/V_{\mathit{NR}}$, $\omega_{\mathit{NR}}$ and $Q_{\mathit{NR}}$ listed in Table 1.       
}
    \label{fig6}
\end{figure}

Figures \ref{fig4}-\ref{fig6} show the central result of this work.  Displayed in the top panel of Fig. 4 (Fig. \ref{fig4}a-c) are measurements of single-tone transmission spectroscopy in the vicinity of $\omega_{cpw}$, versus $\Phi$, for three different values of $V_{\mathit{NR}}$. The corresponding results from numerical simulations, using the parameter values in Table 1, are displayed in Figs. \ref{fig4}d-f.  It is evident from both the data and simulations that for low coupling voltages ($|V_{\mathit{NR}}|\lesssim 5\,{\rm V}$), the cavity response varied with transmon detuning as one would expect from the dispersive interaction between the CPW and transmon.  However, as $V_{\mathit{NR}}$ was increased further ($5\, {\rm V}\lesssim |V_{\mathit{NR}}|\lesssim 7.5\,{\rm V}$), the transmon-nanoresonator interaction became prominent, producing an apparent gap in the CPW transmission spectrum around $\Phi=0.32\Phi_0$, where $\omega_{01} \approx 3.47\, {\rm GHz}$ (Fig. \ref{fig6}), resonant with the simulated value of $\omega_{\mathit{NR}}$. For larger values of coupling ($|V_{\mathit{NR}}|\gtrsim 8\, {\rm V}$, not shown), the cavity response broadened significantly around $0.32\Phi_0$, and the gap was no longer observable.

%%%%% Figure 4

The behavior in Fig. \ref{fig4} can be explained as the interplay between two different effects.  First, as $V_{\mathit{NR}}$ is increased from zero, the corresponding growth of $\lambda$ should lead to hybridization of the nanoresonator and transmon energy levels when $\omega_{01}\approx \omega_{\mathit{NR}}$, producing the well-known phenomenon of Rabi doublets\cite{haroche2006exploring} in the coupled-system's energy spectrum.  Of course, for the temperatures at which these transmission measurements were made, the transmon and nanoresonator should have each resided predominantly in their ground state (transmon and nanoresonator thermal occupancies should both have been $n_{th}\approx 0.004$ at $T=30$ mK), resulting in the joint state $\ket{00}$ and no change in the transmission response of the dispersively-coupled CPW.  However, the increase in $V_{\mathit{NR}}$ was accompanied by heating of the nanoresonator, believed to be a result of dissipation due to leakage current through the silicon substrate.\footnote{This leakage current is believed to have occurred in the DC bias T-filter and CPW, and was observed to increase in a non-linear fashion with $V_{\mathit{NR}}$, suggesting it was related to breakdown in the silicon between the central trace of the T-filter's capacitor and the ground plane.}  Measurements of the leakage current flowing in the DC bias circuitry provided an estimate of the dissipated power on the order of nano-Watts for the range of $V_{\mathit{NR}}$ shown in Fig. \ref{fig4}.\footnote{For voltages larger than $\sim$ 8 V, the current increased dramatically, ultimately leading to observable heating of the dilution refrigerator's sample stage, as measured with standard resistance bridge thermometry} A simple model and COMSOL simulation of the heating effects indicates that this level of dissipation could indeed heat the nanoresonator to a temperature $T_{\mathit{NR}}$ that is out of equilibrium with the transmon, resulting in nonnegligible thermal population of the nanoresonator ($n_{th} > 0.1$; see Supplemental Material). Qualitatively, through the coupling $\lambda$, the thermally excited nanoresonator thus served as an effective thermal bath for the transmon, increasing the probability for the transmon to be found in its first excited state $\ket{m=1}$ at $\Phi\approx0.32\Phi_0$, and leading to a thermally-averaged dispersive shift of the CPW response. 

The physics due to the thermally excited nanoresonator are captured quantitatively in the numerical simulations by increasing $T_{\mathit{NR}}$  simultaneously with $\lambda$.  In Figs. \ref{fig4}d-f, the best-fit by eye to the data was found by increasing $T_{\mathit{NR}}$ from 30 mK for $\lambda/2\pi=1.35 \, {\rm MHz}$ to 180 mK for $\lambda/2\pi=1.95\, {\rm MHz}$, which would have corresponded to an increase in thermal occupation of the nanoresonator mode from $n_{th}=0.004$ to $n_{th}=0.8$.\footnote{It should be noted that for the simulations, $T_{\mathit{NR}}$ serves as the only free-parameter; the remaining parameters (Table 1), were all determined through independent means (such as two-tone spectroscopy or single-tone spectroscopy.} While $n_{th} < 1$ for these values of $T_{\mathit{NR}}$, the numerical simulations show that in the steady-state the increased probability for occupation of the nanoresonator's excited states is enough to enhance the population of the transmon's $m=1$ state and deplete the population of the $m=0$ state (Figs. \ref{fig5}a-d), when the transmon and nanoresonator are near resonance ($\omega_{01}\approx\omega_{NR}$).  This is reflected in numerical calculations of the transmon state probability versus $\Phi$, which show the excited state population increasing directly with $T_{\mathit{NR}}$ and $\lambda$, around $\Phi=0.32\Phi_0$, even while the transmon temperature is held fixed in the simulation at the base temperature of the refrigerator ($T_{Q}=30$ mK).  

It should be noted that, in two-tone spectroscopy measurements of the transmon at $V_{\mathit{NR}}=$ -5 V, the transmon's $\omega_{12}$ transition was not observable above the noise floor of the measurement set up for flux biases where the nanoresonator and transmon were tuned off resonance (Fig. 2b). This indicated that the transmon was not heated directly by the application of $V_{\mathit{NR}}$ and that its enhanced excited ($m=1$) population was in fact due to the thermally excited nanoresonator.  This conclusion is also supported by measurements of the number-state splitting of the transmon's $\omega_{01}$ that arises due to the dispersive interaction with the CPW mode (Fig. 2c). This splitting effect was measured from $V_{\mathit{NR}}=0 \,{\rm{V}}$ to $V_{\mathit{NR}}=-8 \,{\rm{V}}$ and at large detuning in energy from the nanoresonator; the sharp transitions exhibited no observable change in linewidth or peak height as a function of $V_{\mathit{NR}}$, indicating that heating of the transmon (and CPW mode) was negligible over this coupling range.    

It is important to point out that we observed the CPW quality factor to degrade as $|V_{\mathit{NR}}|$ was increased, for values of $\Phi$ where $\omega_{01}\approx\omega_{NR}$. In this range, we observed the CPW linewidth to change from $\kappa_{cpw}/2\pi=0.282\,{\rm{MHz}}$ for $V_{\mathit{NR}}=-4.5\rm V$ to $1.08\,{\rm{MHz}}$ for $V_{\mathit{NR}}=-6.5\rm V$. This increase in $\kappa_{cpw}$ could not be reproduced in the simulations with our model. Thus for the simulation we used the respective measured values of $\kappa_{cpw}$ for $V_{\mathit{NR}}=-4.5\,{\rm {V}}$, $-5.5\,{\rm {V}}$, and $-6.5\,{\rm {V}}$. We believe that this effect could be due to the direct nanoresonator-CPW coupling which we did not include in the model.\footnote{While the CPW-nanoresonator direct coupling is expected to be negligible for the unitary state evolution of our hybrid system, it is not necessarily negligible for dissipative processes affecting the CPW cavity. Because of the relatively high Q of the CPW, the interaction with the lossy nanoresonator may present an important dissipative channel for the CPW features. This remains the subject of future work.}
%%%%% Figure 5

%%%%% Figure 6

The influence of the nanoresonator as a dissipative bath coupled to the transmon in the weak interaction limit explored here (i.e. $\lambda/\kappa_{\mathit{NR}}<<1$) is further substantiated through two-tone spectroscopy measurements of the transmon $\omega_{01}$ transition for frequencies around $\omega_{\mathit{NR}}$ (Fig. 6). These measurements show a clear increase in the linewidth $\gamma$ for $\omega_{01}$ as $\Phi$ was tuned through $\Phi=0.32 \Phi_0$. The origin of the broadening of the transition can be understood through a quantum noise model\cite{schoelkopf2003qubits,clerk2010introduction} - where the transmon-nanoresonator interaction is treated to lowest-order in perturbation theory, yielding a simple relationship between $\gamma$ and the spectral density of the nanoresonator's displacement fluctuations, given by (See Supplemental Material)
\begin{equation}
\label{eq:gamma}
\gamma=\frac{\lambda^2}{x_{zp}^2\hbar^2}(S_x(\omega)+S_x(-\omega))+\gamma_0,
\end{equation}
where $\gamma_0$ represents contributions to transmon dissipation and dephasing that are assumed to be uncorrelated with the nanoresonator's fluctuations and constant over the narrow range of the nanoresonator's response. Here the positive and negative frequency noise components $S_x(\pm \omega)$ are given by the usual relations\cite{clerk2010introduction}
\begin{equation}
\label{eq:posnoise}
S_x(\omega)= x_{zp}^2\frac{\kappa_{\mathit{NR}}(n_{th}+1)}{(\omega_{\mathit{NR}}-\omega)^2+(\frac{\kappa_{\mathit{NR}}}{2})^2}
\end{equation}
and
\begin{equation}
\label{eq:negnoise}
S_x(-\omega)= x_{zp}^2\frac{\kappa_{\mathit{NR}}n_{th}}{(\omega_{\mathit{NR}}+\omega)^2+(\frac{\kappa_{\mathit{NR}}}{2})^2}.
\end{equation}
For small $T_{\mathit{NR}}$ (i.e. $n_{th}\ll1$), such as in Fig. 6,  the nanoresonator should act primarily as a ``cold" bath, preferentially absorbing energy from the transmon. In this limit, Eq. \ref{eq:gamma} reduces to
\begin{equation}
\label{eq:simplegamma}
\gamma=\frac{\lambda^2}{\hbar^2}\frac{\kappa_{\mathit{NR}}}{(\omega_{\mathit{NR}}-\omega)^2+(\frac{\kappa_{\mathit{NR}}}{2})^2}+\gamma_0
\end{equation}
indicating that the frequency dependence of $\gamma$ in this narrow frequency range should be determined by the nanoresonator's susceptibility. Indeed, a fit of Eq. \ref{eq:simplegamma} to the data in Fig. 5b, allowed us to extract $\omega_{\mathit{NR}}/2\pi=3.47$ GHz, $Q_{\mathit{NR}}=\omega_{\mathit{NR}}/\kappa_{\mathit{NR}}=150$, and $\lambda/2\pi V_{\mathit{NR}}= 300$ kHz/V, which we used in the numerical simulations of the single-tone spectroscopy (Fig. 4). Moreover, the observed increase in $\gamma$ around $\omega_{01}=\omega_{NR}$ is consistent with estimates of nanoresonator-induced radiative damping made from a quasi-lumped-element model of the admittance seen by the transmon (See Supplemental Material).    

We recognize that this simple quantum noise model neglects the influence of higher-level transmon states, which we leave as the subject of future work.  Nonetheless, the quantitative agreement between the single-tone spectroscopy data and numerical simulations utilizing these extracted parameter values suggests that the influence of the higher-level states should present minor corrections to this picture.    

\section{Future Prospects and Conclusions}
Based on the results presented in the previous section, we envision three future directions of research with this novel hybrid quantum electromechanical system, which are attainable with realistic improvements to the engineering of the device. First, this new device, operated in the same weak coupling regime ($\lambda/\kappa_{\mathit{NR}}\ll1$) demonstrated here, offers prospects for exploring the quantum noise properties of the nanoresonator.  Eqs. \ref{eq:gamma}-\ref{eq:negnoise} illustrate how the transmon could be used as a spectrometer to resolve the asymmetry in nanoresonator's quantum noise.  With minor modifications to the present device to enable the controlled tuning of $T_{\mathit{NR}}$, the asymmetry between the nanoresonator's positive and negative frequency noise could be carefully mapped through measurements of the transmon's $T_1$ and polarization in the vicinity of $\omega_{\mathit{NR}}$. Such measurements could be implemented over a large range of temperatures (from deep in the quantum regime, $n_{th}\ll 1$, to $n_{th}\sim 10$) - and because they wouldn't require the simultaneous use of sideband techniques to damp and cool the mechanical mode, would thus provide a complimentary approach to recent experiments in circuit optomechanics studying quantum noise of mechanical systems\cite{safavi2012observation,lecocq2015resolving}.  

A second (and related) direction is the use of the nanoresonator as an engineered reservoir to which the transmon could be controllably coupled for exploring the influence of specially tailored thermal and nonthermal baths. Structured baths that differ from the standard Ohmic form and hence display non-Markovian behavior are currently a subject of considerable theoretical interest, particularly when the environment contains some number of strongly coupled discrete modes\cite{thorwart2004dynamics,brito2008dissipative,hausinger2008dissipative,gan2010non,iles2014environmental,levi2016coherent}. Recent experiments have demonstrated the feasibility of characterizing and even actively engineering such non-Markovian environments in optomechanical\cite{groblacher2015observation} and circuit QED systems \cite{haeberlein2015spin}.  Our hybrid system, with the in situ control over component couplings and frequency detuning, would thus be an excellent candidate for further pursuit of such studies, which also can help to pave the way for implementations of controllable meso-nanoscale machines envisioned in the new field of quantum thermodynamics.\cite{millen_perspective_2016} 

As a final direction, we envision the development of this system as a platform to explore quantum coherent dynamics of the coupled CPW mode, nanoresonator and transmon. The degree of tunability of both the transmon's transition energy and the nanoresonator's coupling energy would provide a versatile set-up for exploring both the resonant and dispersive regimes of interaction between the transmon and the nanoresonator and CPW. With realistic improvements to the transmon-nanoresonator coupling strength (discussed below), along with the strong dispersive coupling between CPW and transmon that is already realized in this device, this system could thus be utilized for exploration of fundamental topics related to quantum information and sensing\cite{armour2008probing,utami2008entanglement,asadian2014probing}, as a new hybrid-system for quantum state-generation in two-resonator cQED\cite{abdi2015quantum,mariantoni2008two}, and for further development as an element for implementation in future quantum processing circuits. 

The use of this hybrid quantum device as a platform for exploring coherent dynamics of coupled mechanical and circuit degrees of freedom will require several engineering upgrades to the nanoresonator and transmon that push the two components fully into the strong-coupling regime.   First, improvements can be made to the engineering of the transmon to increase $T_2^\ast$ by at least a factor of 10.  This can be accomplished through two steps: by designing the transmon's ``sweet-spot" in energy, where it is insensitive at first-order to low-frequency flux noise, to be more closely tuned to the nanoresonator frequency than in the present design; and by utilizing an asymmetric junction design, which reduces the transmon's susceptibility to dephasing due to low-frequency flux noise. Such designs should enable $T_1, T_2^\ast$ $\sim 20\,\mu$s in 2D cQED architectures.\cite{ware2015flux}  Second, the coupling strength to the nanoresonator $\lambda$ can be improved by eliminating the heating due to the DC voltage bias and allowing for application of coupling voltages $V_{\mathit{NR}}\gtrsim 6\,{\rm{V}}$. This could be achieved either through the use of a sapphire substrate or thin-film silicon nitride layer in the T-filter, both materials of which have been previously used in voltage-biased qubit-mechanical devices without unwanted heating effects for $V\gtrsim 10\,{\rm{V}}$\cite{lahaye2009nanomechanical,pirkkalainen2013hybrid}. Assuming parameters similar to the current device, the application of $V_{\mathit{NR}}=15$ V, with modest improvement of qubit coherence time to $T_2^\ast$=10 $\mu$s, would yield $\lambda/\gamma\sim 40$, safely within the strong-coupling regime with respect to transmon decoherence.  Finally, truly reaching the strong-coupling limit will require improving the nanoresonator quality factor $Q_{\mathit{NR}}$ by at least a factor of 10.  To accomplish this will require reducing clamping losses, either through the engineering of ``free-free" structures\cite{huang2005vhf}, or the use of phononic band gaps at the supports.\cite{yu2014phononic}  Ultimately, improving the mechanical quality factor could require integrating graphene membranes, which have demonstrated $Q$'s in excess of 100,000 for low-lying flexural modes.\cite{weber2014coupling}

In conclusion, we have demonstrated the operation of a new hybrid quantum system that integrates a high-quality superconducting qubit and microwave circuitry with UHF nanomechanics. We have shown through a comparison of spectroscopic measurements and numerical simulations that the system is well-described by a generalized multi-mode Jaynes-Cummings Hamiltonian, with the strongly-damped nanoresonator serving as a dissipative bath to the qubit.  With realistic improvement to the existing design, we believe this device could soon be compatible with state-of-the-art architecture currently being used in the development of superconducting quantum processors, as well as enable a large range of experiments to study the coherent quantum dynamics and quantum thermodynamics of this complex system.

\section{Acknowledgment}
The authors thank B. Plourde for helpful conversations and technical assistance. This work was performed in part at the Cornell NanoScale Facility, a member of the National Nanotechnology Infrastructure Network, which is supported by the National Science Foundation (Grant No. ECCS-1542081). FB and AOC are supported by Instituto Nacional de Ci\^encia e Tecnologia - Informa\c{c}\~ao Qu\^antica (INCT-IQ) and by Funda\c{c}\~ao de Amparo \`a Pesquisa do Estado de S\~ao Paulo  (FAPESP) under grant number 2012/51589-1. MDL,  FR and Y.H. acknowledge support for this work provided by the National Science Foundation under Grant No. DMR-1056423 and Grant No. DMR-1312421.

\section*{References}

%\bibliography{Rouxinol_HybridQuantumSystem_final}
\putbib[Rouxinol_HybridQuantumSystem_final]
\end{bibunit}
\begin{bibunit}

\clearpage
\setcounter{footnote}{0}
\setcounter{section}{0}
\setcounter{figure}{0}
\setcounter{table}{0}

\section*{Supplemental Material}

%%%%%%%%%%%
%%%%%%%%%%%
\section{Nanoresonator Frequency and Coupling Strength }
\label{coupling}
\subsection{Eigenfrequencies}

The resonance frequencies of the first and third mode of the nanobeam were estimated both analytically and using finite-element numerical simulations. In both cases, because of the small dimensions, it was important to incorporate in the model the native aluminum oxide layer that forms naturally during the fabrication process\cite{Oxidation}. We estimated this oxide surface layer to have a thickness, $t_{AlO_x}\simeq 3$ nm \cite{evertsson_thickness_2015}. Moreover, because the mechanical properties of this AlO$_x$ layer deviate significantly from those of the Al inner core\cite{spearing_materials_2000}, we find that they give rise to significant corrections to the effective density and Young's modulus of the composite structure (Table \ref{table_1S}) for the specific geometrical parameters designed here and thus are essential to include in the calculations of the nanostructure's eigenfrequencies (Table \ref{table_2S}).

In our simulations the nanostructure's dimensions were fixed at 94 nm thick, 38 nm wide and 700 nm long, with an oxide layer of 3 nm on its surface - so that the overall dimensions are consistent with our observations from SEM images of samples fabricated in the same batch as the reported device. For the analytical calculations, the standard expression from continuum mechanics for prismatic thin beams was used to calculated the resonances frequencies of the flexural modes, $f_n$ \cite{Cleland2003}:

\begin{equation}
f_n = 
\frac{k_n w}{l^2}\sqrt{\frac{Y}{\rho}}
\label{eq_freq}
\end{equation}
where $k_n $ accounts for the mode shape, $w$ the width of the beam, $l$ its length, $Y$ the effective Young Modulus and $\rho$ the effective density of the material.
Using the parameters in Table \ref{table_1S}, we estimate the eigenfrequency of the third mode to be $f_3=3.2 \,{\rm{GHz}}$, within the range of the observed value $f_{exp}$, considering typical fabrication tolerances and uncertainties in material parameters: $|(f_3-f_{exp})/f_{exp}|\simeq 8\%$.\footnote{Note that in this section we have used $f_3$ and $f_{exp}$ in place of $\omega_{NR}/2\pi$ and $\omega_{NR,meas}/2\pi$ , which are used in the main text.} 

Numerical calculations of the resonance frequencies of the nanostructure were performed using the finite-element software COMSOL \cite{comsol_multiphysics_2011} (Fig. \ref{fig1S}). Using this program the native oxide layer at the surface of the beam could be incorporated in the calculation of the structure's eigenfrequencies. From these calculations, we found $f_3=3.4 \,{\rm{GHz}}$, yielding a deviation of $|(f_3-f_{exp})/f_{exp}|\simeq 2\%$ from the observed value, well within range of fabrication and materials uncertainties. The mode shapes for both the first and third mode are displayed in Fig. \ref{fig1S}.

\subsection{Coupling Strength}

We assumed the coupling strength between the nanoresonator and the qubit is given by the standard expression \cite{LaHaye2009}:
\begin{equation}
\lambda_n=-4\frac{E_C}{\hbar}\frac{\rmd C_{\mathit{NR}}}{\rmd x}\frac{V_{\mathit{NR}}}{e}x_{zp,n},
\label{eq_lambda}
\end{equation}
where $E_C$ is the electrostatic energy of the qubit, $V_{NR}$ the voltage difference between the transmon and the nanoresonator, $ C_{NR}$ the  capacitance between the transmon and the nanoresonator, and $x_{zp,n}$ represents the zero-point motion of the resonator, given by 
\begin{equation}
 x_{xp,n}= \sqrt{\frac{\hbar}{2m_n\omega_{n}}}
\label{eq_ZPM}
\end{equation}
where $m_n=\alpha_n \rho\, w\, t\, l$ is the effective mass, $\alpha_n$ the mode-dependent effective mass ratio \cite{hauer_general_2013}, $\rho$ the effective density of the beam and $\omega_n$ the resonance frequency of the $n^{th}$ mode\cite{Cleland2003}. Estimates of the zero-point fluctuations for the first mode $x_{zp,1}$ and third mode $x_{zp,3}$, using Eqs. \ref{eq_freq} and \ref{eq_ZPM}, are listed in Table \ref{table_2S}.\footnote{Note that in the main text we have dropped the subscript $n=3$ in the definition of $x_{zp}$ and $\lambda$ for the third mode.}

\begin{table}[]
\centering
\caption{Young modulus $(Y)$ and density $\rho$ for aluminum and aluminum oxide, and the effective density and effective Young modulus of the nanoresonator .}
\label{table_1S}
\begin{tabular}{| c | c | c  | }
\hline\hline
Material   & $\rho$(kg/m$^3$) & $Y$ (GPa)  \\ \hline\hline
Al            &     2700         &       70                    \\ 
Al O$_x$ & 3950         & 380                    \\ \hline
Nanobeam & 2966 & 116 \\ \hline           
\end{tabular}
\end{table}

In order to estimate the coupling strength $\lambda_n$, the capacitance between the nanoresonator and the transmon $C_{NR}(x)$ as a function of their separation $x$  was simulated using finite-element methods by the  software package ANSYS Q3D\cite{_ansys_q3d}.  The simulation was carried out incorporating the silicon substrate, the etched region around the NR, and the native oxide layer. 

For $C_{NR}(x)$ calculated at $x=35 \,{\rm{nm}}$ - the typical transmon-nanoresonator spacing observed from SEM images of other samples fabricated in the same batch as the measured device - a rough estimate for the coupling strength of the third mode $\lambda_{approx,3}/V_{\textit{NR}}$, was derived using the approximation $\case{\rmd C_{NR}}{\rmd x} \simeq \case{C_{NR}}{x}$:

\begin{equation}
\lambda_{approx,3}= -4\frac{E_C}{\hbar}\frac{C_{\mathit{NR}}}{x}\frac{V_{\mathit{NR}}}{e}x_{zp,3}= 195 \,{\rm {kHz}},
\label{eq_lambda_approx}
\end{equation}
within range of the values obtained from spectroscopy measurements presented in the main text ($\sim 300$kHz/V), given the usual uncertainty in the device's geometrical parameters due to fabrication tolerances.     

\begin{figure}[t!]
\centering
\includegraphics[ width=1\columnwidth,keepaspectratio]{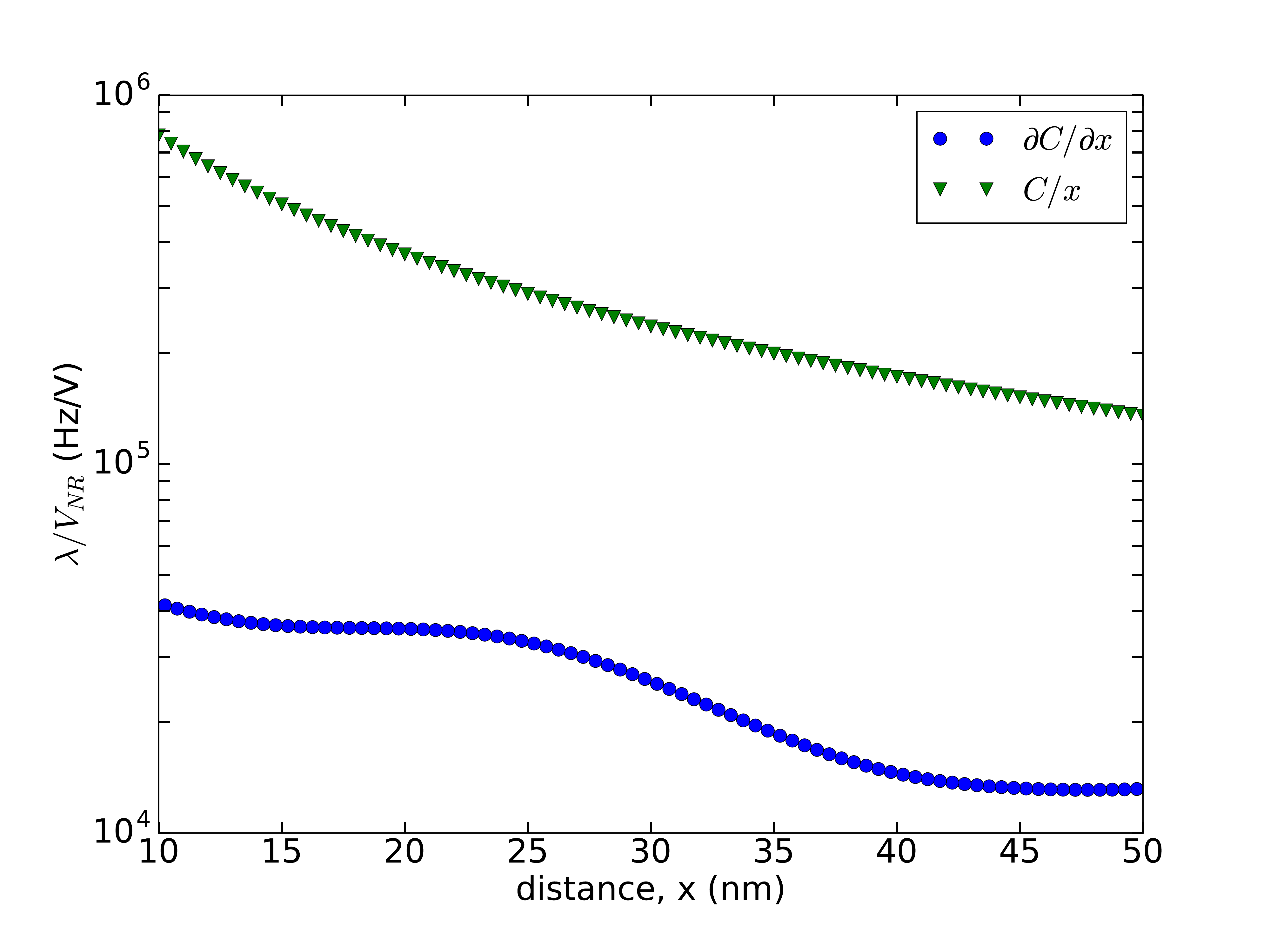}
\caption{ \footnotesize Transmon-nanoresonantor coupling strength plotted versus the separation distance $x$ between the two devices. The calculated coupling strength between the transmon and the nanoresonator for $\case{\rmd C_{\mathit{NR}}}{\rmd x}$ (circles) is derived numerically from $C_{NR}(x)$;  the approximation (triangles) is given by $\case{\rmd C_{\mathit{NR}}(x)}{\rmd x}\simeq\frac{C_{\mathit{NR}}(x)}{x}$
}
\label{fig1S}
\end{figure}

We also estimated the coupling strength $\lambda_3/V_{NR}$ by numerically differentiating $C_{NR}(x)$ around $x=35 \,{\rm{nm}}$ and then using this value in Equation \ref{eq_lambda}. However, we found this estimate of $\rmd C_{NR}(x)/ \rmd x$  to be a factor of $\sim$10 smaller than $C_{NR}(x)/x$ (Fig. \ref{fig1S}), correspondingly yielding a much smaller value of $\lambda_3$  than observed. The divergence between these values is intriguing, and investigations to elucidate its origins are under investigation and will be the subject of future work.  

It should be noted that even though the coupling between the transmon and the fundamental mode of the nanoresonator was approximately 2-3 times greater than the coupling to the third mode, we don't believe the fundamental mode played a significant role in the dynamics of the device, due to the large detuning ($\sim3\,{\rm{GHz}}$) from the transmon (the two systems would have been in a very weak dispersive coupling regime); thus we did not include it in the model used for the simulations that we performed. 

%Table 2S
\begin{table}[t!]
\centering
\caption{Parameters and calculated values for the resonators' first and third modes.}
\label{table_2S}
\begin{tabular}{|c|c|c|}
\hline\hline
Parameter   & n=1 & n=3  \\ \hline\hline
$\alpha_n$ 	& 0.3959 	& 0.4358 \\
$k_n$            &     1.028         &       5.555                    \\ 
$f_n$ (analytically) 		& 590 MHz         & 3200 MHz                   \\ 
$\omega_{NR,n}/2\pi$ (COMSOL) &708 MHz 	& 3400 MHz\\
$x_{zp}$ & 	60 fm		& 25 fm \\
 $\lambda_{approx,n}/2\pi V_{NR}$      &	453 kHz/V		&  195 kHz/V\\ 
  $\lambda_n/ 2\pi V_{NR}$     &	42 kHz/V		&  18 kHz/V\\ \hline     
$E_C/h$         & \multicolumn{2}{c|}{0.227 GHz}                  \\ 
 $C_{NR}/d$         & \multicolumn{2}{c|}{1.4 nF/V ($d$=35nm)}                  \\ \hline
\end{tabular}
\end{table}

\begin{figure}[t!]
\centering
\includegraphics[ width=1\columnwidth,keepaspectratio]{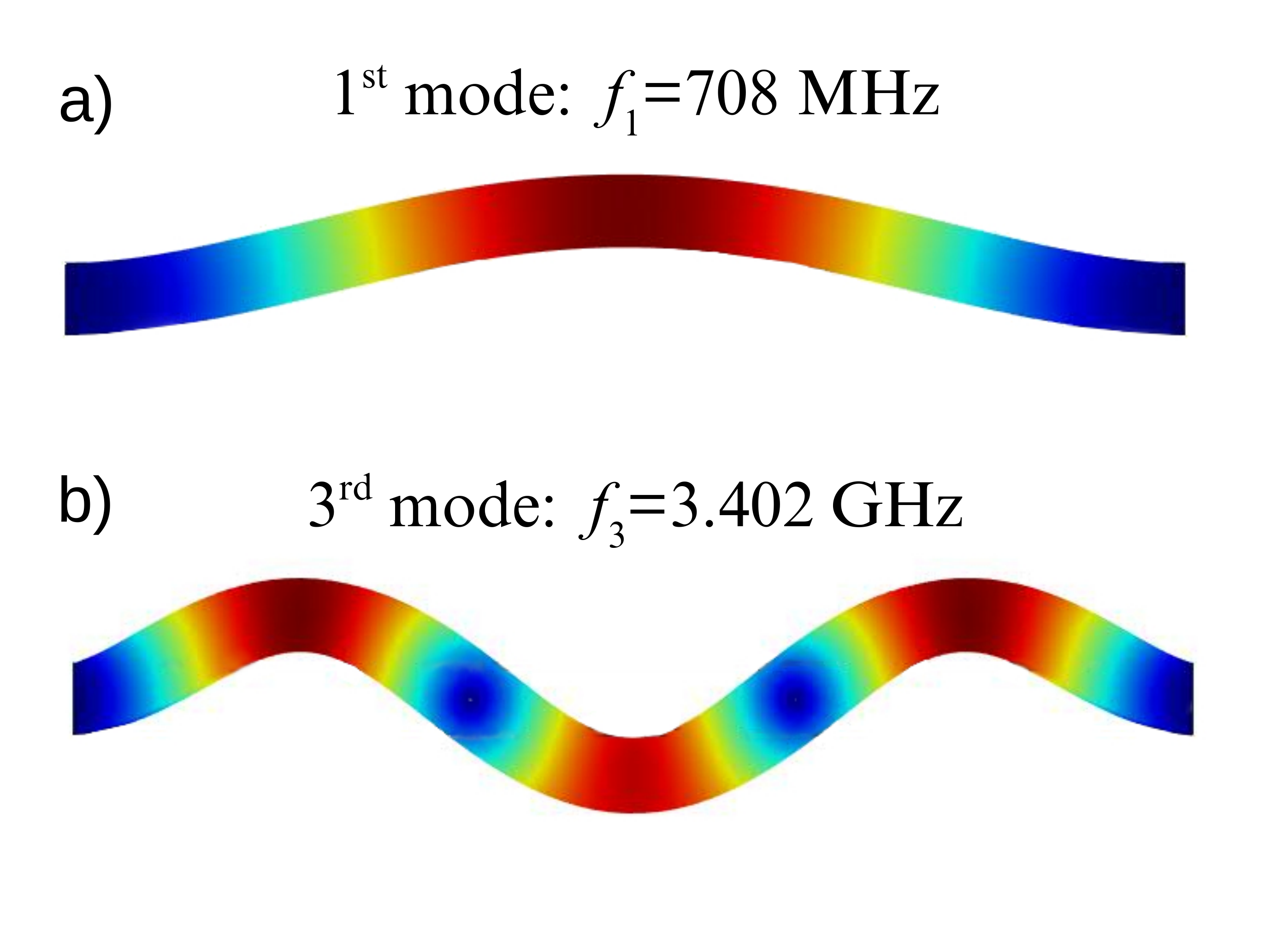}
\caption{ \footnotesize COMSOL simulation of the first and third mode and its eigenfrequencies. 
}
\label{fig2S}
\end{figure}

\section{Heating}

The most convincing explanation for the unwanted heating that we observed of the nanoresonator was leakage current $I_{NR}$ in the sample between the ground plane and the center trace of the CPW and T-filter. We believe that this leakage current was due to a breakdown of the high-resistivity silicon substrate. For pure silicon, the breakdown voltage is approximately $3 \times 10^7$V/m,\cite{sze2006physics} which we believe is of the same order of magnitude as the electric field between the CPW or T-filter center line and ground plane (for the present geometry and $V_{NR} \sim 10 \,{\rm {V}}$), which we estimate to be $\sim 10^7\,{\rm{V/m}}$. 

In order to understand the heating, we measured the current $I_{NR}$ in the nanoresonator's DC bias line when high voltage $V_{NR}$ was applied. This leakage current $I_{NR}$ was observed to vary in a highly nonlinear manner with $V_{NR}$. Below 7 V, $I_{NR}$  was below the resolution of our setup, and on the order of pico-amps or smaller. However, for $V_{NR}=8 \,{\rm{V}}$, $I_{NR}=5 \,{\rm{nA}}$, resulting in $40 \,{\rm{nW}}$ Joule heating on the chip. 

\begin{figure}[t!]
  \centering
    \includegraphics[width=1\columnwidth]{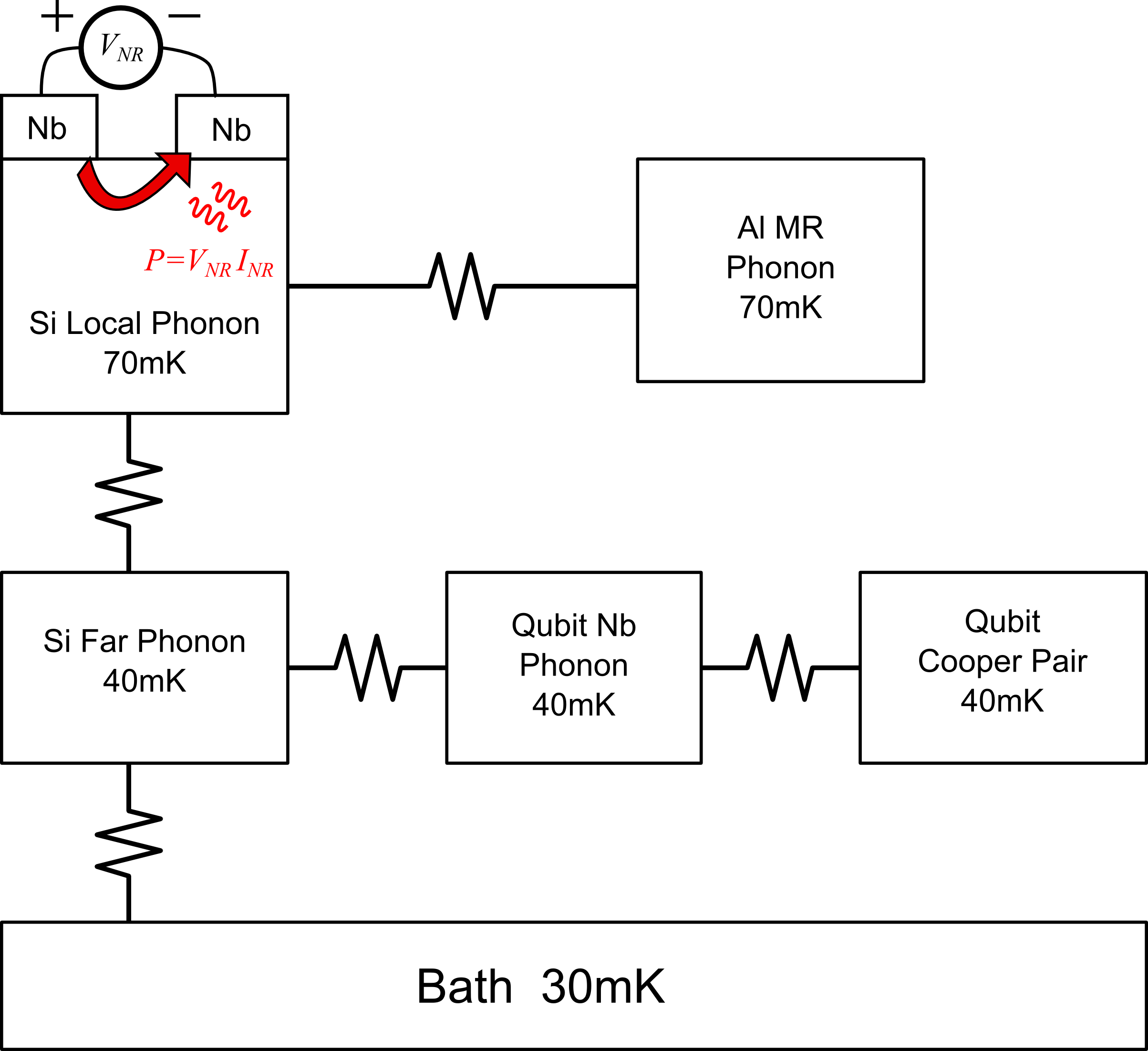}
  \caption{\footnotesize Thermal circuit for modelling the heating due to leakage current in the device. We assume that heat is generated near the surface of the silicon substrate between the Nb center line and Nb ground plane, with dissipated power given by $P=V_{NR}  \cdot I_{NR}$ . This region is denoted as ``Si Local Phonon'' in the figure. Because the nanoresonator is directly connected to center line, and in close proximity to this region, we assume the nanoresonator mode is in equilibrium with the local phonon population.  By contrast, the transmon junctions are geometrically much farther away from where the heat is dissipated (denoted as "Si Far Phonon"), with a smaller thermal resistance to the sample holder, which we assume to be held at the temperature of the refrigerator ($T_{0}= 20 \,{\rm{mK}}$). This results in the electronic degrees of freedom of the junctions being out of equilibrium with the nanoresonator mode.  The thermal impedances are calculated using similar considerations as in Ref. \cite{savin2006thermal}. The temperature at each location denoted in the figure is determined numerically using a COMSOL simulation, which is illustrated in Fig. \ref{fig4S}. }
\label{fig3S}
\end{figure}

\begin{figure}[t!]
  \centering
    \includegraphics[width=1\columnwidth]{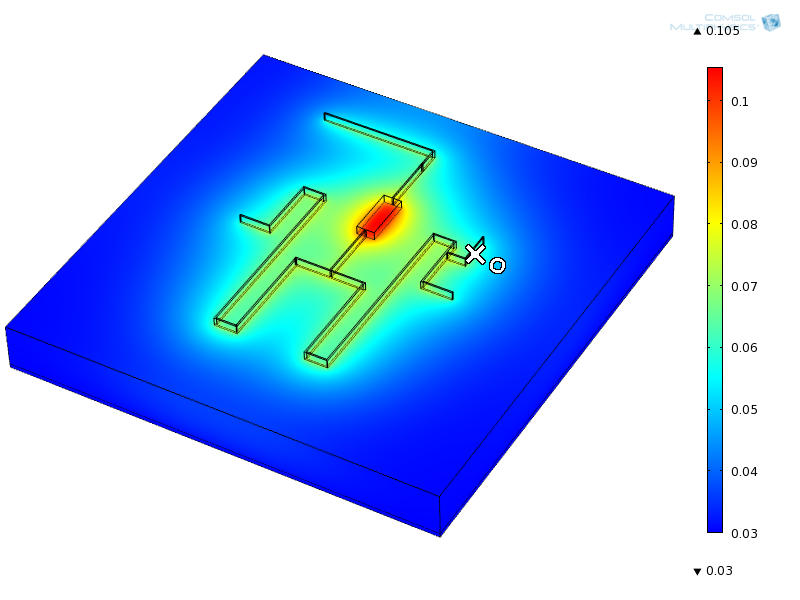}
  \caption{\footnotesize Numerical simulation of the thermal gradient across the device due to the leakage current heating. Color scale indicates temperature in Kelvin as a function of position on the sample.  This simulation was performed by assuming the dissipated power of $40$nW is distributed uniformly along the surface of the silicon between the center trace and ground plane of the CPW and the T-filter. Because the T-filter has a large turn density (due to the inter-digitated capacitor\cite{hao2014development}), much more power is distributed in this region than along the rest of the CPW, resulting in the filter region being heated in excess of 100 mK (darkest red region in the plot).   The $\times$ and small circle $\circ$, denote the respective locations of the nanoresonator and the transmon's junctions.  In the figure, the rectangular slab  represents the $500\mu$m thick silicon substrate. In the simulation, we assume that the bottom of the substrate is connected to the $20$mK refrigerator bath through a $50\mu$m glue layer (too small to see in this view). We are currently developing a more comprehensive model involving the thermal coupling to the electromagnetic environment of the external circuitry.}
\label{fig4S}
\end{figure}

For $V_{NR}\lesssim 8 \,{\rm{V}}$, this dissipated power was too small for noticeable heating of the refrigerator's mixing chamber stage - the fridge stayed around base temperature ($20\sim30\,{\rm{mK}}$). However a simple thermal circuit model (Fig. \ref{fig3S}) indicates that it was large enough to establish a thermal gradient across the sample (Fig. \ref{fig4S}), heating the local phonon modes that couple to the nanoresonator out of equilibrium with the transmon and CPW. 

For simplicity, to model the heating we assumed that the dissipation was generated at the surface of the silicon substrate between the center line and ground plane. Moreover we assumed that the sample-stage of the refrigerator, and the sample holder to which the device was glued, remained in equilibrium at the base temperature of the refrigerator. Using estimates of the various thermal impedances (given in Fig. \ref{fig3S}) of the circuit from standard approaches found in the literature\cite{savin2006thermal}, we then utilized COMSOL to perform numerical simulations of the phonon temperature distribution within the silicon substrate.  

Results from the simulation show that for 40 nW of dissipation distributed evenly along the length of the CPW and T-filter, the nanoresonator temperature could be heated by at least 30 mK above the local phonon temperature in the region of the transmon junctions. In fact the model indicates (not shown) that the temparture gradient could be much larger (and closer to what we observe in the spectroscopy data presented in the main text), depending on the specific values of the thermal impedances and how the heat dissipation is actually distributed along the CPW and filter transmission lines, both of which we still need to investigate in much greater detail.  While it remains a work in progress, we feel this rudimentary model captures the essence of the heating problem, and we are currently using it as a guide to eliminate the problem in the design of the next generation of devices.    

\section{Radiation loss of qubit}

In this section we provide an estimate of the radiative (circuit) damping of the transmon due to its coupling to the T-filtered CPW and the voltage-biased nanoresonator. When the transmon is detuned in energy from the nanoresonator, we find that the relaxation time of the transmon $T_{1}$ is dominated by radiative losses to the T-filtered CPW, and our estimate is in close agreement with the maximum value of $T_1$ we observe in time-domain measurements (Fig. 2a, main text).  Moreover, when the transmon is tuned near resonance with the nanoresonator, our estimates of the increase in linewidth $\gamma$ due to the resonant coupling with the nanoresontor are consistent with our observations from two-tone spectroscopy measurements (Fig. 6, main text).    

In order to estimate the effect of circuit damping on the transmon due to the CPW, T-filter, and nanoresonator, we perform calculations similar to those in the work of Houck \textit{et al}\cite{houck2008controlling}, wherein it is shown that the real part of the admittance seen by the qubit provides an accurate estimate of the qubit's $T_1$. 

To calculate the admittance presented by the nanoresonator, CPW and T-filter, we used a quasi-lumped-element circuit model: the voltage-biased nanoresonator was represented by a lumped-element, series RLC circuit;\cite{truitt2007efficient} the CPW cavity was treated as a transmission line resonator,\cite{pozar2005microwave} terminated with symmetric input/output capacitors, $C_k$, connecting it to the 50 $\Omega$ input and output transmission lines; and the T-filter was represented as a lumped-element inductor-capacitor network connected to the mid-point of the cavity.\cite{hao2014development} In the model, we assumed the transmon was located near one of the voltage anti-nodes of the CPW's fundamental mode. We also assumed a transmon shunt capacitance $C_B=90 \textrm{fF}$, coupling capacitance to the CPW $C_c=10\textrm{fF}$, and $C_k=10 \textrm{fF}$, consistent with the properties of the device reported here.  From this model, we derived an analytical expression for the admittance $Y$ seen by the transmon, which enabled us then to calculate $T_1(\omega_{01})=C_B\rm{Re}[1/Y(\omega_{01})]$ as a function of transmon transition frequency $\omega_{01}$ (Fig. S5).

\begin{figure}[t!]
	\centering
		\includegraphics[width=1\columnwidth]{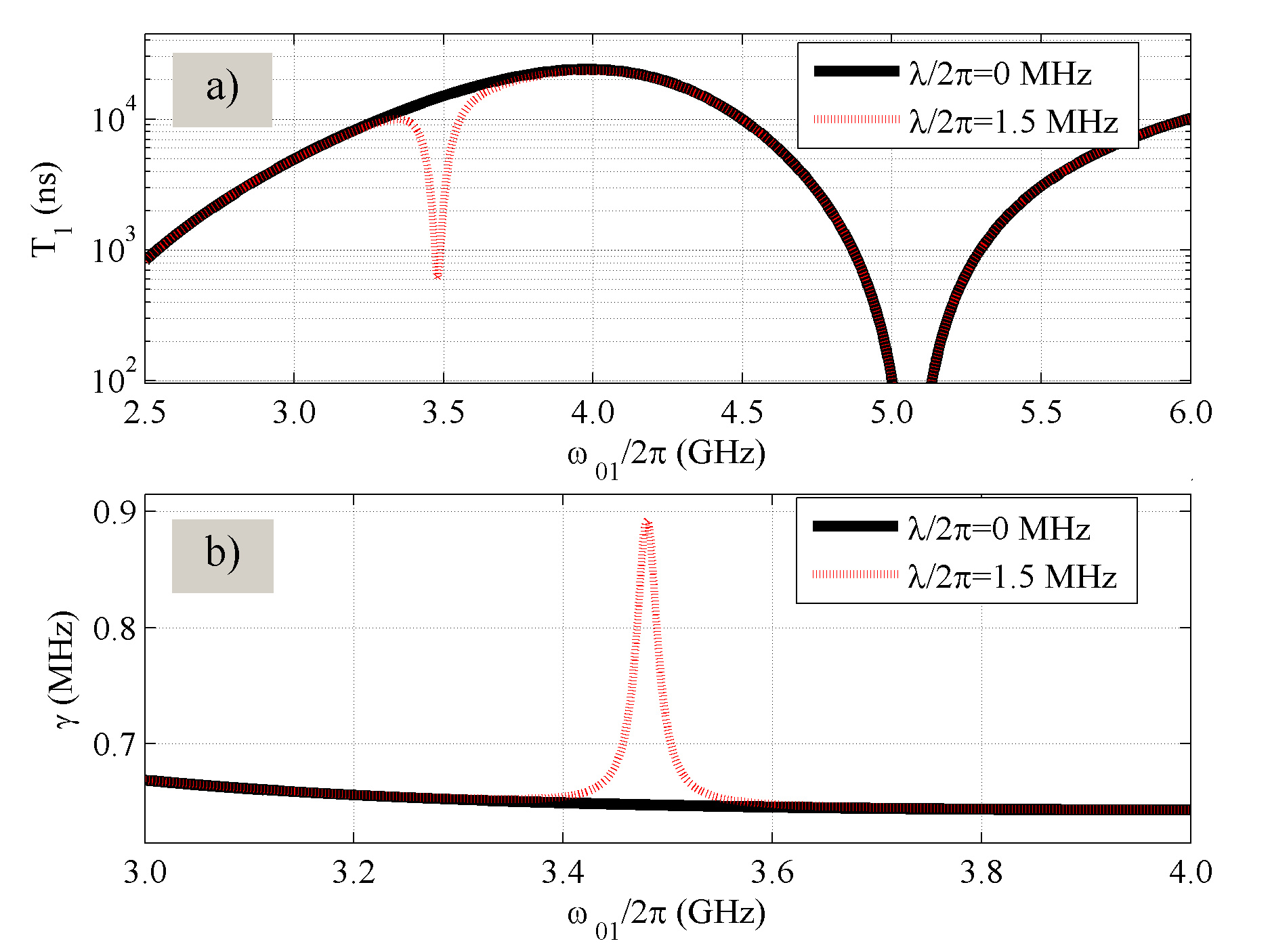}
			\label{fig:radationloss}
	\caption{\footnotesize Estimated relaxation time $T_1$ and linewidth $\gamma$ plotted versus $\omega_{01}/2\pi$ for a transmon qubit capacitively coupled to a voltage-biased nanoresonator, and embedded in a filtered CPW cavity. In (a), $T_1$ is calculated from the admittance presented to the transmon by the nanoresonator, CPW and T-filter. (b) $\gamma$ is calculated from $T_1$ presented in (a) and background dephasing $T_{\phi}$, which is determined from measurements of the transmon linewidth in the vicinity of $\omega_{01}/2\pi=3.5\,{\rm{GHz}}$. In both (a) and (b), the solid balck line is calculated for transmon-nanoresonator coupling $\lambda/2\pi=0.0\,{\rm{MHz}}$; the dashed red line is for $\lambda/2\pi=1.5\,{\rm{MHz}}$, corresponding to $V_{NR}=5\,{\rm{V}}$.}
\end{figure}

Figure S5a shows the expected $T_1$ due do circuit loss as a function of $\omega_{01}$. It is evident that $T_1$  is strongly influenced by the fundamental mode of the cavity near $\omega_{01}= 5\,{\rm{GHz}}$ and by the low-frequency cut-off of the T-filter near $\omega_{01}/2\pi=$ 2 GHz\cite{hao2014development}. Around $\omega_{01}/2\pi=4\,{\rm{GHz}}$, we see that an upper-bound of $T_1 \approx 20 \,{\mu \rm{s}}$ is expected, within range of the measured value of $T_1$ reported in the main text.  Also evident is the increased damping (decreased $T_1$) near $\omega_{01}/2\pi=3.5\,{\rm{GHz}}$, where the transmon and nanoresonator are resonant. Figure S5b displays the estimated linewidth $\gamma$ of the 0-1 transition, for two values of nanoresonator-transmon coupling $\lambda$. It was calculated using $\gamma=1/\pi T_2$, where $1/T_2=1/2T_{1}+1/T_{\phi}$, with $T_{\phi}$ estimated from the background linewidth in Fig. 6b (main text).  For $\lambda/2\pi=1.5\,{\rm{MHz}}$ ($V_{NR}=5\,{\rm{V}}$), the estimated increase in linewidth at $3.5\,{\rm{GHz}}$ ($\sim 250\,{\rm{kHz}}$) is in close agreement with the increase determined from a fit of the two-tone spectroscopy measurements in Fig. 6b ($\sim 280\,{\rm{kHz}}$).

%%%%%%%%%
%%%%%%
\section{Simulation }
\subsection{Model and Numerical Simulations}

To model the interactions between the transmon, nanoresonator and CPW cavity in our device, we performed numerical simulations of the single-tone spectroscopy,\footnote{We did not perform simulations of the two-tone spectroscopy of the complete device, due to the complexity of such simulations, and leave this for future work.} which were based on a multi-level, generalized Jaynes-Cummings Hamiltonian model of the system\cite{pirkkalainen2013hybrid,suri_observation_2013}:

\begin{equation}
\label{eq:totalH}
\hat{H}=\hat{H}_{T}+\hat{H}_{cpw}+\hat{H}_{T-cpw}+\hat{H}_{NR}+\hat{H}_{T-NR}+\hat{H}_{Drive},
\end{equation}
where the transmon's bare Hamiltonian is given by

\begin{equation}
\hat{H}_{T}=\sum_{m}\hbar\omega_{0m}\ket{m}\bra{m},
\end{equation}
with $\omega_{0m}=m\sqrt{8E_{J}(\Phi)E_{C}}/\hbar$ in the limit that $E_{J}(\Phi)/E_{C}\rightarrow\infty$, the qubit's $\ket{0}\to \ket{m}$ transition resonance frequency;

\begin{equation}
\hat{H}_{cpw}=\hbar\omega_{cpw}\hat{a}^{\dagger}\hat{a},
\end{equation}
is the bare Hamiltonian for the CPW mode, with $\omega_{CPW}$ the bare cavity frequency, $\hat{a}^{\dagger} (\hat{a})$ the creation (annihilation) operator for the cavity mode;

\begin{equation}
\hat{H}_{T-cpw}=\sum_{l,m}g_{l,m}\ket{l}\bra{m}(\hat{a}^{\dagger}+\hat{a}),
\end{equation}
is the interaction term between the transmon and CPW, with $g_{l,m}=g\bra{l}\hat{n}\ket{m}$  the coupling strength of the $\ket{l} \to  \ket{m}$ transition, where $\hat{n}$ represents the Cooper-pair number operator;

\begin{equation}
\hat{H}_{NR}=\hbar\omega_{NR}\hat{b}^{\dagger}\hat{b},
\end{equation}
is the bare Hamiltonian for the nanoresonator, with $\omega_{NR}$ the bare nanoresonator frequency,  $\hat{b}^{\dagger} (\hat{b})$ the creation (annihilation) operator for the mechanical mode;

\begin{equation}
\hat{H}_{T-NR}\sum_{l,m}\lambda_{l,m}\ket{l}\bra{m}(\hat{b}^{\dagger}+\hat{b}),
\end{equation}
is the coupling Hamiltonian for the transmon and mechanical mode, with $\lambda_{l,m}=\lambda\bra{l}\hat{n}\ket{m}$; and

\begin{equation}
\hat{H}_{Drive}=E_{d}(e^{i\omega t}\hat{a}+e^{-i\omega t}\hat{a}^{\dagger})
\end{equation}
accounts for the microwaves applied to the cavity to perform single-tone spectroscopy, where $\omega$ is the frequency, and $E_{d}$ is the amplitude of the signal. Note that direct coupling between the nanoresonator and CPW was found to have a negligible influence on the simulations (for the parameter regime obtained experimentally) and thus is omitted in Eq. \ref{eq:totalH}.

Transforming the total Hamiltonian in the rotating frame of the driving term the explicit time dependence can be removed. This was accomplished by

\begin{equation}
\hat{H}_t = \hat{U}\hat{H}\hat{U}^\dagger + i\frac{ \partial \hat{U}}{\partial t}\hat{ U}^\dagger
\end{equation}
using the unitary transformation given by

\begin{equation}
\hat{U} = e^{(i\omega a^\dagger a t)}
\end{equation}
The resulting time-independent Hamiltonian $ \hat{H}_t $ was then truncated to the three lowest levels of the qubit ($\ket{g}, \, \ket{e}$ and  $\ket{f}$); the  energy  levels of the \textit{qutrit} were found from the eigenenergies of the Cooper-pair-box Hamiltonian, using 51 charge states. 

The dynamics of the system, including the interaction of each component with the environment, was modelled via a standard approach using a Lindblad master equation:
\begin{eqnarray}
\dot{\rho}=&&
-\frac{i}{\hbar}[\hat{H},\rho] +
\kappa_{cpw}^- \mathcal{D}[a]\rho +
\kappa_{cpw}^+ \mathcal{D}[a^\dagger]\rho+
\Gamma_{01}^{-} \mathcal{D} [\varrho_{01}^-]\rho\nonumber\\
&&+\Gamma_{12}^{-} \mathcal{D} [\varrho_{12}^-]\rho +
\Gamma_{01}^{+} \mathcal{D} [\varrho_{01}^+]\rho +
\Gamma_{12}^{+} \mathcal{D} [\varrho_{12}^-]\rho +\\
&&\frac{\gamma_{1}^{\varphi}}{2} \mathcal{D} [\varrho_{1}^z]\rho +
\frac{\gamma_{2}^{\varphi}}{2} \mathcal{D} [\varrho_{2}^z]\rho +
\kappa_{NR}^- \mathcal{D}[b]\rho +
\kappa_{NR}^+ \mathcal{D}[b^\dagger]\rho,\nonumber
\end{eqnarray}
where $\mathcal{D}$ is the usual collapse superoperator $\mathcal{D[\mathcal{A}]}\rho\equiv\mathcal{A}\rho \mathcal{A} -\left( \mathcal{A}^\dagger\mathcal{A}\rho+\rho\mathcal{A}^\dagger\mathcal{A} \right)/2$. 

The main parameters of the Hamiltonian and the interaction with the environment were determined experimentally. The CPW resonance frequency and its loaded quality factor, $Q_l$, were determined for each measured voltage  in the vicinity of the nanoresonator frequency ($\omega_{NR}/2\pi\simeq$ 3.47GHz ), see Table \ref{tab:Supp}. The cavity damping rate is represented by $\kappa_{cpw}^-=\kappa_{cpw}$, being estimated from $Q_l$ for each $V_{NR}$ by
\begin{equation}
\kappa_{cpw} =\frac{\omega_{cpw}}{2\pi Q_l}.
\end{equation}
On the other hand, the cavity thermal excitation process is represented by the rate $\kappa_{cpw}^+$, which is approximately determined by
\begin{equation}
\frac{\kappa_{cpw}^+}{\kappa_{cpw}}\simeq \exp{\left(-\frac{\hbar\omega_{cpw}}{k_BT_{cpw}}\right)}
\end{equation}

Relaxation and dephasing rate were determined by time-domain measurements. The temperature $T_{cpw}$ was estimated from an analysis of the black-body radiation incident on the CPW from external circuitry (See Section S6).

The transition between the states $\ket{i}\to\ket{j}$ of the transmon is represented by the operator $\varrho_{ij}\equiv\ket{i}\bra{j}$, and its relaxation and thermal excitation rates by $\Gamma_{ij}^{-}$  and $\Gamma_{ij}^{+}$. The rate of thermal excitation due to the interaction with the environment $\Gamma_{ij}$ can be described by:
\begin{equation}
\frac{\Gamma_{ij}^+}{\Gamma_{ij}^-}\simeq \exp{\left(-\frac{\hbar\omega_{ij}}{k_BT_{Q}}\right)}
\end{equation}
Here $T_Q$ is the transmon temperature, which we estimate an upper-bound on from spectroscopy measurements (See Section S6). For the $\ket{1}\to\ket{0}$ transition, they can be associated with the measured relaxation and decoherence times by

\begin{equation}
T_1 = \frac{1}{\Gamma_{01}^+ + \Gamma_{01}^-}
\end{equation}
and 
\begin{equation}
\frac{1}{T_2^*}=\frac{1}{2T_1}+\frac{1}{T_{\varphi }}
\end{equation}
The dephasing rate of each one of the states ($\varrho_{i}^z\equiv\ket{i}\bra{i}$) is described by  $\gamma_{i}^{\varphi}$. 
The nanomechanical resonator interaction with the environment is described by the rate of loss of phonons $\kappa_{NR}^-=\kappa_{NR}$  and the rate of creation of phonons, due to thermal excitations, $\kappa_{NR}^+$ given by 
\begin{equation}
\frac{\kappa_{NR}^+}{\kappa_{NR}}\simeq \exp{\left(-\frac{\hbar\omega_{NR}}{k_BT_{NR}}\right)}
\end{equation}
From the transmon spectroscopy measurements, shown in Figure 2c of the main text, and the dressed resonant frequency of the CPW cavity, the coupling between the transmon and the CPW was estimated; the values are shown in table \ref{tab:Supp}.

Steady-state solutions $\dot{\rho}=0$ of the master equation were carried out numerically using the open source software package QuTip\cite{qutip} to solve the master equation. The microwave cavity and the mechanical resonator  were simulated using 4 and 5 states respectively. 

\begin{table}[t!]
	\centering 
	\caption{Experimental parameters used to perform the simulations. The coupling between the states of the \textit{qutrit} to the resonator, and the relaxation factor of the CPW in the vicinity of the nanoresonator eigenfrenquency, for different $V_{NR}$ applied to the system.  }
	\begin{tabular}{| c | c |}
		\hline\hline
		Parameter                                       & Value                        \\
		[1ex]
		\hline 
		$g_{01}/2\pi$                                        & 120 $\mathrm{MHz}$           \\
		$g_{12}/2\pi$										&	180 MHz\\
		$\lambda/(2\pi V_{NR})$                         & $\approx 300 \mathrm{kHz/V}$ \\
		$\kappa_{cpw}/2\pi$ (V$_{NR}$=4.5V)             & 0.28MHz                      \\
		$\kappa_{cpw}/2\pi$ (V$_{NR}$=5.5V)             & 0.37MHz                      \\
		$\kappa_{cpw}/2\pi$ (V$_{NR}$=6.5V)             & 1.08MHz                      \\
		$\kappa_{NR}/2\pi$                              & 23MHz                        \\[1ex] \hline
	\end{tabular}
	\label{tab:Supp}
\end{table}
%%%%%%%
\section{Quantum noise of nanoresonator}
In this section we present a simple quantum noise model to understand the broadened qubit linewidth for transition frequencies near the nanoresonator's resonance. In this model we only consider the transmon's lowest two states, treating it as a two-level quantum system that couples to the nanoresonator through an interaction of the form $\hat{H}_{int}=\frac{\lambda}{x_{zp}} \hat{x}(t) \hat{\sigma}_x$ , where $\lambda$ is the coupling strength between the nanoresonator and qubit, $x_{zp}$ is the nanoresonator's zero-point motion defined in Section \ref{coupling} of the Supplementary Material, and $\hat{x}(t)$ represents the nanoresonator's position degree of freedom as a function of time $t$.

For sufficiently small $\lambda$, the interaction can be treated using first-order time-dependent perturbation theory\cite{schoelkopf2003qubits}, which we expect should yield the following relations for the nanoresonator-induced decay and excitation of the qubit:

\begin{equation}
\Gamma_{\uparrow,NR} =\frac{\lambda^2}{x_{zp}^2\hbar^2} S_x(-\omega_{01})\\
\Gamma_{\downarrow,NR} =\frac{\lambda^2}{x_{zp}^2\hbar^2} S_x(\omega_{01})
\end{equation}
where $\omega_{01}$ is the transition frequency of the qubit, and $S_x(\pm \omega)$ is the nanoresonator's displacement spectral density  which can be related to the imaginary part of its response function, $\chi_{x}^{''}(\omega)$, by the well-known fluctuation-dissipation theorem as 
\begin{equation}\label{FDtheorem}
S_x(\omega)=2 \hbar \langle n(\omega) + 1 \rangle \chi_{x}^{''}(\omega)\\
\end{equation}
where
\begin{equation}
\chi_{x}^{''}(\omega)=\frac{1}{m} \frac{\kappa_{NR}\,\omega}{(\omega_{NR}^{2}-\omega^{2})^2 + 4\omega^{2} (\kappa_{NR}/2)^2} \,\, ,\\
\end{equation}
$\kappa_{NR}$ is the nanoresonator's linewidth, $\omega_{NR}$ is the resonance frequency of the nanoresonator (as defined in the main text) and $n(\omega)$ is the thermal occupation number as a function of the frequency.
Using that $\chi_{x}^{''}(\omega)$ is an odd function of its argument and the explicit form of $n(\omega)$,  we can further write
\begin{equation}
S_x(-\omega)=2 \hbar \langle n(\omega) \rangle \chi_{x}^{''}(\omega).\\
\end{equation}
Therefore, if $\omega \approx \pm \omega_{NR}$, we can easily show that
\begin{equation}
S_x(\omega)=x_{zp}^2 \frac{\kappa_{NR}\langle n+1 \rangle}{(\omega_{NR}-\omega)^2+(\kappa_{NR}/2)^2}\,\,,\\
\end{equation}
and
\begin{equation}
S_x(-\omega)=x_{zp}^2 \frac{\kappa_{NR}\langle n \rangle}{(\omega_{NR}+\omega)^2+(\kappa_{NR}/2)^2}\,.\\
\end{equation}  
Assuming that other sources of relaxation for the transmon (i.e. radiative losses, dielectric and interface loss, quasi-particles, etc), which we denote $\Gamma_B$, are uncorrelated with the motional degrees of freedom of the nanoresonator, we can then define the total qubit linewidth as 

\begin{eqnarray}
\gamma(\omega_{01})&=&\Gamma_{\uparrow,NR}+ \Gamma_{\downarrow,NR}+\Gamma_B\nonumber\\
&=&\frac{\lambda^2}{x_{zp}^2\hbar^2} (S_x(-\omega_{01})+S_x(\omega_{01}))+\Gamma_B
\end{eqnarray}
If we now assume that $\omega_{01}$ is an independent variable that we can tune over the narrow range of frequencies around the NR resonance $\omega_{NR}$ (which we can do in experiment by tuning the flux applied to the split-junction qubit), and we assume moreover that $\Gamma_B$ is independent of frequency over this narrow range (which should be accurate given that $\Gamma_{NR} \ll \omega_{01}$), then, in the limit that $\langle n \rangle \ll 1$, the qubit linewidth can be written in simplified form
\begin{equation}
\label{eq:lorentzgamma}
\gamma(\omega) = \frac{\lambda^2}{\hbar^2} \frac{\kappa_{NR}}{(\omega_{NR}-\omega)^2+(\kappa_{NR}/2)^2}+\Gamma_B
\end{equation}
From fits to the data in Fig. 6 of the main text, we estimate $\frac{\gamma(\omega_{NR})-\Gamma_B}{2 \pi} \approx 280 \textrm{kHz}$ and $\frac{\kappa_{NR}}{2 \pi} \approx 24 \textrm{MHz}$. With these parameters determined, one can then use Eq.~\ref{eq:lorentzgamma} to calculate the coupling strength 
\[\frac{\lambda}{h}=\sqrt{\frac{(\gamma(\omega_{NR})-\Gamma_B) \kappa_{NR}}{4(2\pi)^2}} \approx 1.3 \textrm{MHz}\]
For $V_{NR}=5\,{\rm{V}}$ (the value at which data in Fig. 6 of the main text was taken), this yields $\displaystyle{\lambda}/{h V_{{NR}}} \approx 270 \textrm{kHz/V}$, which is very close to the value used in the single-tone spectroscopy simulations displayed in Fig. 4 of the main text.

\section{Transmon Temperature $T_Q$ Estimate}

\begin{figure}[t!]
\centering
\includegraphics[width=1\columnwidth]{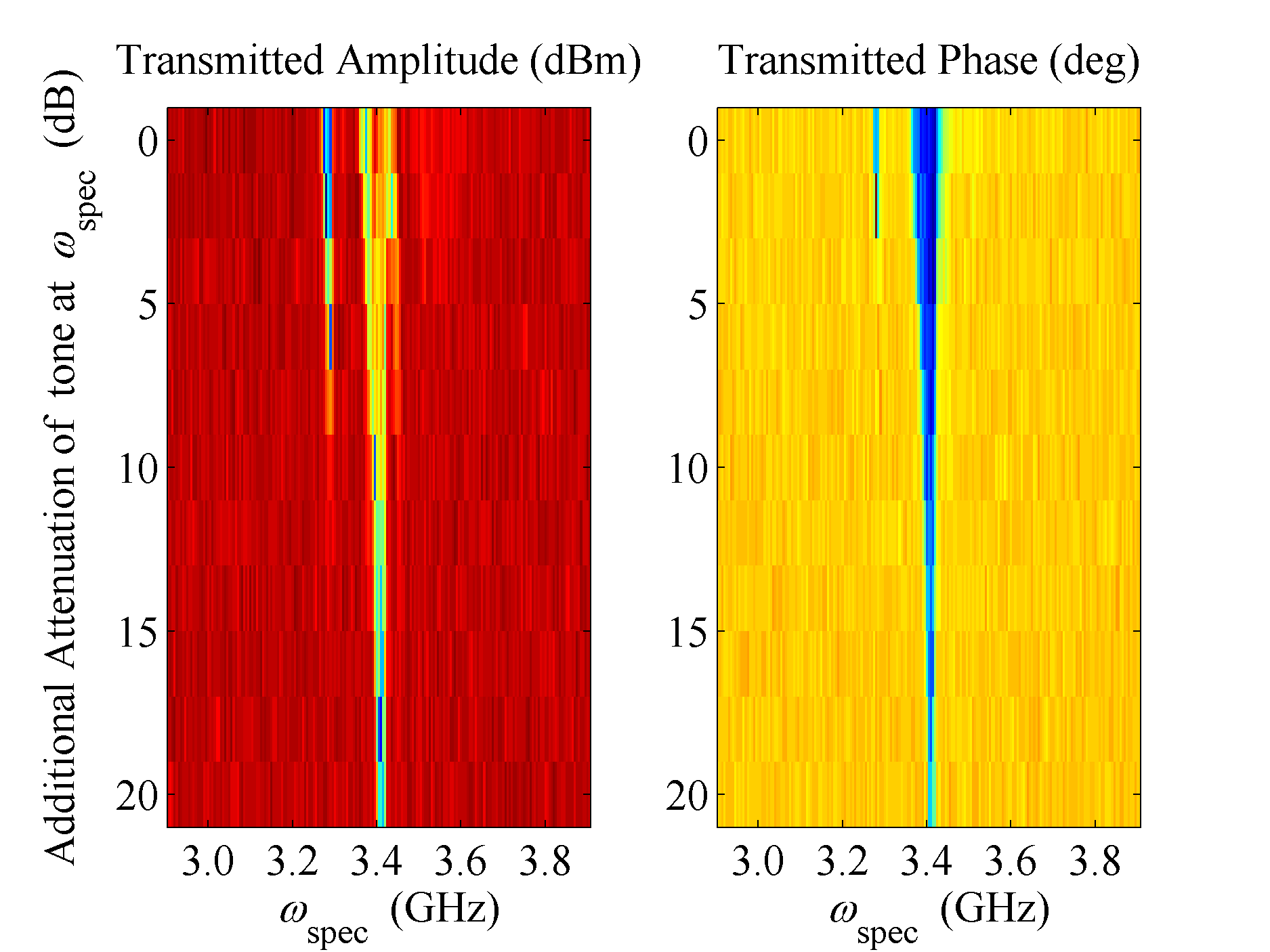}
\caption{\footnotesize Two-tone spectrum of the transmon for $V_{NR}=5\,{\rm{V}}$ versus microwave power (additional external attenuation) and microwave tone $\omega_{spec}$. Flux $\Phi$ is fixed so that $\omega_{01}=3.4\textrm{GHz}$. The left(right) figure is the amplitude(phase) of CPW's transmitted signal. The power of microwaves incident on the transmon is controlled by tuning the additional attenuation (\textit{y} axes) via a digital attenuator located in the input circuitry at room temperature.}
\label{fig:specmap}
\end{figure}

To estimate the effective temperature of the transmon during the experiments reported in the main text, a map (See Fig~\ref{fig:specmap}) of two-tone spectroscopy versus microwave power and spectroscopy tone $\omega_{spec}$ was taken at $\omega_{01}=3.4\textrm{GHz}$, so that $\omega_{01}$, $\omega_{12}$, $\omega_{02}/2$ are all lower than the nanoresonator frequency $\omega_{NR}=3.47\textrm{ GHz}$.

When power was sufficiently low (for additional external attenuation $\gtrsim10$ dB), only the $\ket{0}\to\ket{1}$ transition at $\omega_{spec}= \omega_{01}$ was visible (also see red curve in Fig. 2b in the main text). For lower attenuation, the two-photon $\ket{0}\to\ket{2}$ transition at $\omega_{02}$ became observable at $\omega_{spec}=\omega_{02}/2$.  However, over the full range of applied microwave power, no obvious peak at $\omega_{12}$ was visible above the noise floor of the measurement, indicating a low thermal occupation of the transmon's $m=1$ state.  In order to determine an upper limit on the temperature of the transmon, we estimate the fractional probability of $m=1$ and $m=0$ using the ratio of the noise floor in the vicinity of $\omega_{12}$ (expected to be found at $\omega_{spec}=\omega_{02}-\omega_{01}$) to the peak height of the 0-1 transition at low microwave power ($A_{01}=110\,{\rm{deg}}$ from Fig. 2b of the main text). From the standard deviation in the noise floor $\sigma$ around $\omega_{spec}=\omega_{12}$, we can place a conservative upper limit on the qubit temperature $T_Q$ by estimating $A_{12}<3\sigma=3\times1.85\,\textrm{deg}$. The upper limit on $T_Q$ can then be determined from the Boltzmann factor $\displaystyle \case{A_{12}}{A_{01}}=e^{-\hbar\omega_{01}/{k_B T_{Q}}}$, yielding $T_{Q}< 55\,\textrm{mK}$. For the simulations presented in the main text, we set $T_Q=30\,{\rm{mK}}$.

%%%%%%%%
%%%%%%%%

\section{CPW Fundamental Mode Temperature $T_{cpw}$ Estimate}
In this section we estimate the fundamental mode temperature of the CPW cavity, which we believe to be limited by black-body radiation incident on the cavity from external circuitry.  To estimate the cavity temperature $T_{cpw}$, we assume that the CPW is over-coupled to the input and output transmission lines, which is the case for our measurements ($Q_{cpw,max}\sim 20\times10^3 \ll Q_{intrinsic}\sim10^5$)\cite{hao2014development}. With this assumption we can then estimate $T_{cpw}$ by simply considering the contributions from thermal radiation incident at cavity's input and output ports.  From detailed balance, one finds
\begin{equation}
\kappa_{cpw}n_{cpw}=\kappa_{in}n_{in}+\kappa_{out}n_{out}
\label{kappacpw}
\end{equation}
where $\kappa_{cpw}=\kappa_{in}+\kappa_{out}$ is the total damping rate of the cavity, and $\kappa_{in}$ ($\kappa_{out}$) is the damping rate due to coupling through the CPW’s input (output) port, $n_{in}$ ($n_{out}$) is the incident population of photons at the CPW mode frequency $\omega_{cpw}$ at the input (output) port of the CPW, and $n_{cpw}$ is the thermal population of the CPW mode (i.e. inside the cavity). 

For the case of our device, $\kappa_{in}=\kappa_{out}$, thus $\kappa_{cpw}=2\kappa_{in}$, and so Eq. \ref{kappacpw} becomes
\begin{equation}
n_{cpw}=(n_{in}+n_{out})/2
\label{ncpw}
\end{equation}
In order to calculate the thermal populations $n_{in}$ and $n_{out}$, we assume that they are determined by the thermal radiation from resistive components at higher temperature stages of the refrigerator. For the input line (Fig. 3, main text),  one can see that this includes $50\,\Omega$ resistive attenuators at room temperature, 1 K (30 dB), 700 mK (6 dB), 100 mK (10 dB) and 30 mK (20 dB).  Thus, properly taking into account the attenuation of thermal radiation from higher temperature stages, the input incident photon population is estimated to be 
\begin{equation}
n_{in}=\frac{n_{300\,{\rm{K}}}}{10^{6.6}}+\frac{n_{1\,{\rm{K}}}}{10^{3.6}}+\frac{n_{700\,{\rm{mK}}}}{10^3}+\frac{n_{100\,{\rm{mK}}}}{10^2}+n_{30\,{\rm{mK}}}
\label{nin}
\end{equation}
Using the Bose-Einstein occupation factor to calculate each of the contributions in Eq. \ref{nin}, we find that $n_{in}=0.0050$.  

A similar calculation can be made for $n_{out}$, except on the output line we assume that the thermal radiation is dominated by the $50\,\Omega$ input resistance of the HEMT amplifier thermalized at 4 K (Fig. 3, main text).  This amplifier is connected to the output port of the CPW through a section of superconducting coaxial cable and two cryogenic circulators in series that are thermalized to the mixing chamber of the refrigerator.  Properly taking into account the $D=30-35$ dB of isolation provided by the two isolators leads to the following expression for $n_{out}$:
\begin{equation}   
 n_{out}=\frac{n_{4\,{\rm {mK}}}}{10^{D/10}}+n_{30\,{\rm{mK}}}
\label{nout}
\end{equation}
which yields an incident photon population of $n_{out}=0.0055-0.0085$. 

Using Eqs. \ref{ncpw}-\ref{nout}, we find an effective cavity mode temperature of $T_{cpw}=40-50\,{\rm{mK}}$. For the single-tone spectroscopy simulations presented in Fig. 4 of the main text, we thus set $T_{cpw}= 45\,{\rm{mK}}$. We also used this value of $T_{cpw}$ for the simulation plot of the number-state splitting that is presented in Fig. 2c (main text).  

\section*{References}
\putbib
%\bibliography{Rouxinol_HybridQuantumSystem_final}
\end{bibunit}
\end{document}